\documentclass[12pt]{article}
\usepackage{amsmath,amsfonts,amssymb,amsthm}

\usepackage{graphicx}

\if@twoside  m
\else
\fi \topmargin 5mm \headheight 0mm \headsep 0mm \textheight
225truemm \textwidth 150truemm
\newcommand{\Real}{{\mathbb {R}}}           
\newcommand{\Com}{{\mathbb {C}}}            
\newcommand{\Nat}{{\mathbb {N}}}            
\newcommand{\Int}{{\mathbb{Z}}}            

\newcommand{\C}{{C\!\!\!\rule[.5pt]{.7pt}{6.5pt}\:\:}}
\newcommand{\nid}{\noindent}
\newcommand{\vsth}{\vskip2mm}

\def\bra{{\langle}}
\def\ket{{\rangle}}

\def\eps{\varepsilon}
\def\erf{{\rm erf\,}}

\newcommand{\dist}{{\rm dist\,}}

\newcommand{\id}{{\rm id\,}}

\newcommand{\Ran}{{\rm Ran\,}}

\newcommand{\BB}{{\cal B}}
\newcommand{\OO}{{\cal O}}

\newcommand{\EE}{{\cal E}}

\newcommand{\HH}{{\cal H}}
\def\OO{{\cal O}}
\def\RR{{\cal R}}
\numberwithin{equation}{section}

\begin{document}
 
\title{ The dynamics of 1D Bloch electrons in constant electric fields }
\author{F. Bentosela${}^{a)}$, P. Duclos${}^{a)}$, G.  Nenciu${}^{b)}$ and V. Moldoveanu${}^{c)}$}
\maketitle
\vspace{0.5cm}
${}^{a)}${\it Centre de Physique
  Th\'{e}orique UMR 6207 - Unit\'e Mixte de Recherche du CNRS et des
Universit\'es Aix-Marseille I, Aix-Marseille II 
et de l' Universit\'e
du Sud Toulon-Var - Laboratoire affili\'e \`a la FRUMAM, F-13288
Marseille Cedex 9, France}
\vspace{0.5cm}

${}^{b)}${\it Department of Theoretical Physics, University of Bucharest, P.O. MG 11, 76900 - Bucharest, Romania 
and Institute of Mathematics of the Romanian Academy, P.O. Box 1-764, RO70700 Bucharest, Romania. }
\vspace{0.5cm}

${}^{c)}${\it Group of Theoretical Physics, National Institute of Materials Physics,
P.O.Box MG-7, Bucharest, Romania.}

\begin{quote}

\noindent {\small
{\bf Abstract.} We study the dynamics of a 1D Bloch electron
subjected to a constant electric field.
The periodic potential is supposed to be less singular 
than the $\delta $-like potential (Dirac comb). We give a rigorous proof
of Ao's result \cite{Ao} that for a large class of initial conditions 
(high momentum regime) there is no localization in momentum space. The proof is based 
on the mathematical substantiation of the two simplifying assumptions 
made in physical literature: the transitions between far away bands 
can be neglected and the transitions at the quasi-crossing can be described 
by Landau-Zener like formulae. 
Using the connection between the above model and the driven quantum ring (DQR)
shown by Avron and Nemirovski \cite{AvN}, our results imply the increase of energy
for weakly singular such DQR and appropiate initial conditions. 
  }   
\end{quote}
 
 
\section{Introduction}
\par
The dynamics of Bloch electrons (i.e electrons subjected to a periodic potential)
 in the presence of a homogeneous electric field is among those topics that appeared 
since the 
beginning of the quantum theory of the solid state physics but
are still alive today for both  mathematical and physical reasons; in particular
the emergence of superlattices leads to interesting physics (see e.g \cite{V})

The dynamics of an electron in one dimension subjected to a periodic potential
$V_{{\rm per }}(x)$ and a constant electric field $E=-eF$ is described by 
the time-dependent Schr\"{o}dinger 
equation:
\begin{equation}
i\hbar\frac{\partial\psi }{\partial t}(x,t) =H^{SW}\psi (x,t)
=\left (-\frac{\hbar ^2}{2m}\frac{d^2}{dx^2} -eFx+V_{{\rm per}}(x)\right )\psi (x,t) 
\end{equation} 
where 
  $H^{SW}$ is the  so-called 
Stark-Wannier Hamiltonian.

There is a large body of mathematically oriented literature about spectral 
properties of the Stark-Wannier Hamiltonian (especially concerning the  
 Stark-Wannier ladder problem; see \cite{Bentosela2,Bentosela3,Bentosela4,
n1} and references therein). In particular the spectrum has been proved to be 
absolutely continuous for twice differentiable potentials \cite{BCDSSW} and recently,
results about the nature of the spectrum were obtained 
for more singular potentials \cite{ki1,k2,P,ADE}.

On the contrary, at the rigorous level there are by far fewer results 
concerning the dynamics generated by $H^{SW}$. Of course, as well known, the results
about spectral properties of $H^{SW}$ lead to results about dynamics but they 
either concern finite (albeit long) intervals of time \cite{n1,n2} or are only qualitative. 
As an example we mention the results of Avron and Nemirovski \cite{AvN} on driven quantum rings. 
They first proved the nice result that there is a close connection between
driven quantum rings and Stark-Wannier Hamiltonians. Then, using this connection
and the fact mentioned above that the spectrum of $H^{SW}$ is absolutely continuous
for twice differentiable potentials they proved that the energy of a smoothly 
driven quantum ring grows indefinitely as $t\to\infty $ for arbitrary initial 
conditions. 

At the physical level the dynamics generated by $H^{SW}$ has been thoroughly studied
(using the temporal gauge representation) both by       
analytical and numerical methods. Due to the difficulty of the problem 
two main assumptions are made:

1. The only interband transitions considered are the ones
between neighbouring bands ( the so-called Zener tunneling process through 
avoided crossings); all the others are considered sufficiently small 
to be neglected.

2. The transition probabilities can be computed approximatively 
via Landau-Zener type formulae.

Based on earlier developements \cite{Br}, the consequences of 
these assumptions were exhaustively discussed by Ao \cite{Ao}.  
More precisely, writing the evolution at arbitrary times 
as a product of evolutions over half Bloch periods and using for the last ones 
the "scattering matrix" between adjacent bands given by Blatter and Browne \cite{Br}
he reduced the original problem to a discrete dynamics, amenable to
numerical and analytical study. The main result coming from 
his analysis is that for potentials more regular than the Dirac comb 
and a large class of initial conditions there is a propagating front for the Bloch electron
(in other words it will escape at infinity). The Dirac comb is a critical
border: one could have either propagation or localization (respectively pure point, 
continuous spectrum 
spectrum or even mixt cases ) depending upon  
 the electric field strength and some 
resonance 
conditions. 

At the rigorous level the existence of a propagating front 
(with quantitative estimates) for a large class of initial conditions 
as well as some spectral consequences has been recently proved in \cite{ABDN}
for potentials more regular than the $L^2_{{\rm loc}}$ class. 

The aim of this paper is to provide a rigorous justification of the
simplifying assumptions 1 and 2 above in order to substantiate from 
the mathematical point of view at least a part of Ao's analysis.
We do it for a class of periodic potentials whose Fourier
coefficients, ${\hat V}(n)$, satisfy  $|{\hat V}(n)|<{\rm const.}|n|^{-r}$
 for all $r>0$. Notice however that the results of sections 3 and 4 are even
valid for $r>-\frac{1}{2}$.

The need for a rigorous control comes from the fact that even if the 
errors involved in  the assumptions 1 and 2 are "small" over a 
half Bloch period they acummulate during a long time evolution to the point 
of making irrelevant the approximate computation based on the discrete
dynamics.

The problem turned out to be fairly complex
for two reasons. First we are dealing with singular periodic potentials 
which leads to a definition in the form sense of the gauged Hamiltonian ${\tilde H}^{SW}$
 and secondly, we want to control the evolution over infinite intervals of time which 
demands a very good control of the errors involved. 
  
The results in this paper allows us to prove the existence of a  propagating front
(which in turn implies the existence of continuous spectrum for $H^{SW}$ ) up to the 
Dirac comb (which corresponds to ${\hat V}(n)=1$ and is known as notoriously 
difficult). Actually the Dirac comb case remains open, although it is not
clear for us whether the present approach could provide as well some results
in this case.   

The plan of the paper is as follows. Section 2 contains the preliminaries: the direct 
integral representation of the Stark-Wannier Hamiltonian in the temporal gauge 
, the periodicity in time properties of the fiber Hamiltonian and the reduction 
to the one Bloch period. All that is done for ${\tilde H}^{SW}$ defined as a quadratic
form sum. Section 2 contains in addition a key technical estimate (Lemma 2.3)
used many times in the following sections. Since the proof is technical and somehow 
long it was moved to Appendix 7.1.  

Section 3 contains the first main result of our paper (Theorem 3.1) saying that the 
transitions over a half Bloch period, between far-away bands (neglected 
by assumption 1 in the physical literature ) can be controlled by  adiabatic techniques.
The idea is that in adiabatic perturbation theory the relevant parameter
is $\varepsilon / \Delta $, where $\varepsilon $ is the slowness parameter and
$\Delta $ is the spectral gap. In our case it is the gap that grows with energy
making thus the adiabatic machinery effective. 

After the far-away decoupling in Section 3 we come to the ``in-band" dynamics
i.e to Zener transitions through quasi-crossings at high energies. The task of computing 
this dynamics has been taken again and again in the physical literature. 
It consists in solving a
$2\times 2$ system of first order ordinary differential equations over a finite 
interval of time. There are two problems here. The first is that the coefficients 
are taken as given by a low order almost degenerate pertubation theory 
without any control of the reminder. Secondly, the obtained ``Zener model" is solved 
over an infinite period of time. We cope with the first problem by using 
the reduction theory in Section 4 to obtain an effective Hamiltonian 
(Theorem 4.1). Then we integrate the system over one half period by the use of Dyson 
series; we compute the first two terms in Section 5 and estimate the remainder 
in Appendix 7.3. The resulting "transfer matrix" is given in Theorem 5.1. 
We expect the results mentioned above to give a lot of information on 
spectral and dynamical properties of the Stark-Wannier (SW) and driven 
quantum ring models for fairly singular potentials. 
As an example in Section 6 the results in the previous sections are assembled 
to prove the Ao's statement about the the existence of a  propagating front. 
As direct side consequences we obtain (see Corollary 6.4) that for $r>0$ 
the continuous spectrum of $H^{SW}$ is not empty (see however the recent paper
of Perelman \cite{P} for a better result in this direction), that 
for a large class of initial conditions the energy of the DQR increases 
like $t^2$ (see Corollary 6.5) and that there is no localization in momentum space 
(see Corollary 6.6). 

In Appendix 7.2 we state for convenience the Sz-Nag\"y transformation matrix. 
In order to make the reading of the paper
easier we start each section by simply stating the theorems and lemmae, their proofs
being provided later.

\newpage
\setcounter{equation}{0}
\section{Preliminaries}                                     

\par 
As already  said in the introduction we are interested in the evolution given by the 
Stark-Wannier operator. For simplicity we normalize $e=\hbar=2m=1$. 
Moreover since the results we are going to prove along the paper do not depend upon 
the  electric field strength, we set $F=1$. Accordingly the Stark-Wannier Hamiltonian writes
\begin{equation}\label{2.1}
H^{SW}=-\frac{d^2}{dx^2}-x+V_{{\rm per }},
\end{equation}
where $V_{{\rm per }}$ is a real periodic potential with periodicity $a=2\pi$.
 Under these assumptions the quantity $\frac{2\pi\hbar}{aF}$, the so-called Bloch period, is 
equal to 1.
\\
Let us define:
\begin{equation}\label{2.2} 
{\tilde U}^{SW}(t,s):=G(t)e^{-i(t-s)H^{SW}}G(s)^*,
\end{equation} 
where $G(t)$ is the temporal gauge: 
\begin{equation}\label{tg}
G(t)\psi (x,t):=e^{-ixt}\psi (x,t)=:{\tilde\psi}(x,t).
\end{equation}
Then using (\ref{2.2}) and (\ref{tg}) one has by direct computation:
\begin{eqnarray*}
i\frac{d}{dt}{\tilde U}^{SW}(t,s)&=&
i\left (\frac{d}{dt}G(t)\right )G(t)^*{\tilde U}^{SW}(t,s)+G(t)H^{SW}e^{-i(t-s)H^{SW}}G(s)^*\\
&=&\left (i \left (\frac{d}{dt}G(t)\right )G(t)^*+G(t)H^{SW}G(t)^*   \right ){\tilde U}^{SW}(t,s)\\
&=&\left ( \left ( -i\frac{d}{dx}+t\right )^2+V_{{\rm per}}(x)
       \right ){\tilde U}^{SW}(t,s).
\end{eqnarray*}
Denoting 
\begin{equation}
{\tilde H}^{SW}(t):=  \left ( -i\frac{d}{dx}+t\right )^2+V_{{\rm per}}(x)
\end{equation}
we get 
\begin{equation}
i\frac{d}{dt}{\tilde U}^{SW}(t,s)={\tilde H}^{SW}(t){\tilde U}^{SW}(t,s).
\end{equation}
As well known the periodicity of $V_{{\rm per}}$ allows a direct integral representation 
for ${\tilde H}^{SW}(t)$ via the  Fourier-Bloch transform \cite {RS4} which maps unitarily the space 
$L^2 (\Real)$
onto $L^2\left ([0,1),dk,l^2({\Int})\right ) $:
\begin{equation}\label{2.3}
(S\psi )(k,n)=\frac{1}{\sqrt {2\pi}}\int_0^{2\pi}e^{-inx}\left\lbrace\sum_{\gamma\in\Int }
e^{-ik(x+2\pi\gamma )}\psi (x+2\pi\gamma) \right\rbrace dx.
\end{equation} 
Note that $(S\psi )(k,n)=({\cal F}\psi)(k+n)$ where ${\cal F}$ is the Fourier transform.
In this representation ${\tilde H}^{SW}(t)$ writes  
\begin{equation}\label{2.4}
S{\tilde H}^{SW}(t)S^*=\int ^{\oplus}_{[0,1)} H(k,t)dk,
\end{equation}
with the fiber Hamiltonian
\begin{equation}\label{2.5}
H(k,t):=H_0(k,t)+V.
\end{equation}
We remark that $H(0,t)$ coincide with the Hamiltonian of the DQR when written in the
 Fourier representation (see e.g \cite{AvN}) so all the results below on the evolution 
$U(t,s)$, generated by $H(0,t)$ apply (see the Corollary 6.5 below) to the DQR problem.  
The unperturbed Hamiltonian $H_0(k,t)$ has a simple spectral
representation
\begin{equation}\label{2.6}
H_0(k,t)=\sum_{n\in \Int}E_{n,0}(k,t)P_{n,0}(k),
\end{equation}
where
\begin{equation}\label{2.7}
E_{n,0}(k,t)=(n+k+t)^2
\end{equation}
and $P_{n,0}(k):=P_{n,0}$ is the projection on the $n^{th}$ vector, 
 $\varphi _n $, from the canonical basis
in $l^2(\Int)$ 
\begin{equation}\label{2.8}
\varphi_{n,0}(m)=\delta_{n,m}.
\end{equation}
The perturbation $V$ is given by the following convolution in $l^2(\Int)$ 
\begin{equation}\label{2.9}
( V\phi) (k,m)=\sum_{n\in \Int}{\hat V}(m-n)\phi (k,n),
\end{equation}
 ${\hat V}(n)$ being the Fourier coefficients of $V_{{\rm per }}$. 
Since $V_{{\rm per }}$ is real we have ${\hat V}(n)=\overline {\hat V}(-n)$.
Moreover we can choose $\hat V(0)=0$ (this amounts for 
 a shift in the energy scale).
We characterize different classes of perturbations by the norm
\begin{equation}\label{2.10}
\|V\|_r:=\sup _{n\in \Int}\langle n\rangle ^r|\hat V(n)|<\infty,
\qquad  \langle n\rangle:=\sqrt {1+n^2}
\qquad r\in\Real .
\end{equation}
Notice that when $r$ gets smaller and smaller, $V_{{\rm per }}(x)$ is more and more singular.
Actually $V_{{\rm per }}(x)$ is to be considered as   
 a tempered distribution on the one
dimensional torus $\mathbb{T}$ such that its Fourier series satisfies $\|V\|_r<\infty$. For $r>0$, 
$V_{{\rm per }}(x)$ is realized as an usual function 
while $r=0$ corresponds to 
$\delta$-like potentials. For $r>\frac{1}{2}$, $V_{{\rm per }}(x)$ 
is square integrable over the unit cell,
and then as well known \cite{RS4}, $V$ is $H_0(k,t)$- bounded with relative bound zero.
As a consequence, by the Kato-Rellich theorem \cite{RS4}, $H(k,t)$ is self-adjoint on the 
(time independent) domain of $H_0(k,t)$: 
\begin{equation}
{\cal D}(H_0(k,t))={\cal H}^2=\{\phi(n)| \langle n\rangle ^2\phi(n)\in l^2({\mathbb{Z}})\}.
\end{equation}
In addition, the existence of a strongly continuous evolution $U(k;t,s)$, generated by $H(k,t)$
is assured by standard results (see  Thm. X.70 in  \cite{RS2}). 
We are mainly interested in the case of more singular potentials corresponding to $r\leq\frac{1}{2}$ 
and here the problem is more involved since as one can easily see (take for example 
${\hat V}(n)=\langle n \rangle ^{-1/2}$ ) $V$ is no more $H_0(k,t)$-bounded. Notice that the 
time independent form domain of
$H_0(k,t)$ is ${\cal Q}(H_0(k,t))={\cal H}^1=\{\phi(n)| \langle n\rangle\phi(n)\in l^2({\mathbb{Z}})\}$. Then one can use the theory of Hamiltonians defined as quadratic forms    
 \cite{RS2}. More precisely, from 
\begin{equation}
R_0(z,k,t):=\sum_{n\in\Int }\frac{P_{n,0}}{E_{n,0}(k,t)-z}
\end{equation}
let us define
\begin{equation}
|R_0(z,k,t)|^{\frac{1}{2}}=\sum_{n\in\Int }
\frac{P_{n,0}}{|E_{n,0}(k,t)-z|^{\frac{1}{2}}}
\end{equation}
and 
\begin{equation}
R_0(z,k,t)^{\frac{1}{2}}=
\sum_{n\in\Int }\frac{ {\rm sgn}(E_{n,0}(k,t)-z)P_{n,0}}{|E_{n,0}(k,t)-z|^{\frac{1}{2}}}.
\end{equation}
Then 
$R_0(z,k,t)=R_0(z,k,t)^{\frac{1}{2}}|R_0(z,k,t)|^{\frac{1}{2}}$. 
The following operator,
defined for $z\notin \sigma (H_0(k,t))$ will appear many times along the paper:
\begin{equation}\label{2.14}
K(z,k,t):=|R_0(z,k,t)|^{\frac{1}{2}}VR_0(z,k,t)^{\frac{1}{2}}
\end{equation}
The fact that $V$ is $H_0(k,t)$-bounded in the form sense 
with relative bound zero is implied by the following lemma:
\par
\vskip 0.2cm
\noindent {\sl {\bf Lemma 2.1 :} Let $r>-\frac{1}{2}$.
Then uniformly in $t$ and $k$
\begin{equation}\label{2.15}
\lim _{a\to\infty} \|K(-a^2,k,t)\|=0.
\end{equation}
}
As a consequence one can use the KLMN theorem \cite{RS2} 
to define $H(k,t)$ as a form sum with form domain
${\cal H}^1$. The associated quadratic form is
\begin{equation}\label{2.16}
\HH^{1}\ni \phi\to\sum_{n}|(n+k+t)\phi(k,n)|^2+\sum_{m,n\in\Int}{\hat V}(m-n)
\overline{\phi}(k,m)\phi(k,n).
\end{equation}
Due to Lemma 2.1 one can write a useful formula for $R(z,k,t):=(H(k,t)-z)^{-1}$. Indeed, for 
sufficiently large $a$    
\begin{equation}\label{2.17}
R(-a^2,k,t)=R_0(-a^2,k,t)^{\frac{1}{2}}(1+K(-a^2,k,t))^{-1}|R_0(-a^2,k,t)|^{\frac{1}{2}},
\end{equation}
one has by analytic continuation
\begin{equation}\label{2.17a}
R(z,k,t)=R_0(z,k,t)^{\frac{1}{2}}(1+K(z,k,t))^{-1}|R_0(z,k,t)|^{\frac{1}{2}}.
\end{equation} 
Since $|R_0(z,k,t)|^{\frac{1}{2}}$ is compact, it follows from (\ref{2.17}) that $R(z,k,t)$
is compact which implies (see \cite{RS4}) that the spectrum of $H(k,t)$ is discrete. 
In what concerns 
the evolution $U(k,s,t)$ generated by $H(k,t)$, as a corollary of 
Thm. 2.27 in \cite{S} one has the 
following result assuring the existence of $U(k;s,t)$ in the weak sense:

\vskip 0.2cm
\noindent {\sl {\bf Lemma 2.2}  Let $\HH^{-1} $ denote the space of continuous 
linear forms 
on $\HH^1$. 
Then for every  $\phi(k,0)\in\HH^1$ it exists a unique function
 $\phi (k,\cdot) :\Real\to\HH^1$ such that
\vsth
\nid (i)
 $\Real\ni t\to (\phi(k,t),\psi)$ is continuous for 
any $\psi\in\HH^{-1}$,
\vskip1mm
\nid (ii)
for every $f\in\HH^1$,
$$
{\frac{1}{i}}\frac{d}{dt}(\phi(k,t),f)+(\phi(k,t),H(k,t)f)=0.
$$
\vskip1mm
\nid (iii) The map $U(k,t,s):\phi(k,s)\to\phi(k,t)$ is isometric and its extension by continuity is unitary 
in $l^2(\Int)$.
}
\vsth
\nid 
From (\ref{2.2}), (\ref{2.4}) and  Lemma 2.2 we can write the following formula for the 
Stark-Wannier evolution 
\begin{equation}\label{2.18}
e^{-i(t-s)H^{SW}}=G(t)^*S^*\int ^{\oplus}_{[0,1)} U(k,t,s)dk\,SG(s),
\end{equation}
so the study of $e^{-i(t-s)H^{SW}}$ is reduced to the study of the family of evolutions $U(k,t,s)$
in $l^2(\Int )$ generated by $H(k,t)$, and the rest of the paper will be devoted to this problem.  
From (\ref{2.5}) - (\ref{2.9}) it follows that $H(k,t)=H(0,k+t)$ and then
\begin{equation}\label{2.19}
 U(k,t,s)=U(0,k+t,k+s),
\end{equation}
which means that one can restrict the analysis to the fiber $k=0$, and from 
now on we omit to write $k$
when it is taken to be zero e.g. $H(k=0,t)=H(t)$, $E_{n,0}(k=0,t)=E_{n,0}(t)$ etc. The next remark is 
that although $H(t)$ is not periodic in $t$ there is a hidden periodicity. More precisely if $T$
is the shift operator in $l^2(\Int )$:
\begin{equation}\label{2.20}
(T\psi )(n)=\psi (n-1)
\end{equation} 
then by direct computation one can see that
\begin{equation}\label{2.21}
TH(k,t+1)T^*=H(k,t),
\end{equation}
which in turn implies
\begin{equation}\label{2.22}
U(t+1,s+1)=T^*U(t,s)T,
\end{equation}
The analysis of $U(t,s)$ can be restricted to one period. We take the basic period to be the union 
of two half-periods:
\begin{equation}\label{2.23}
t\in I_0\cup I_1; 
\qquad I_0= [-\frac{1}{4},\frac{1}{4} ),  
I_1=\left [\right. \frac{1}{4},\frac{3}{4}\left.\right ).   
\end{equation}
Let us consider now in more detail the spectral properties of $H_0(t)$ and $H(t)$. We remark that 
when $t\in I_0\cup I_1 $, $H_0(t)$ has degeneracies at the points $t=0$ and $t=\frac{1}{2}$. 
Actually the splitting of one period into two  half-periods is due to the existence of 
two points per period where
$H_0(t)$ has degeneracies. Since the latter play a key role in the dynamics (due to the Zener transitions)
the above structure suggests to describe $\sigma (H_0(t))$ inside $I_l$, $l=0,1$ as a union of pairs 
of eigenvalues that cross at $t=0$ and $t=\frac{1}{2}$. We define (see Fig.1)

\begin{figure}
\begin{center}
\vspace*{-6cm}
\hspace*{1cm}
      \includegraphics[angle=00,width=16cm]{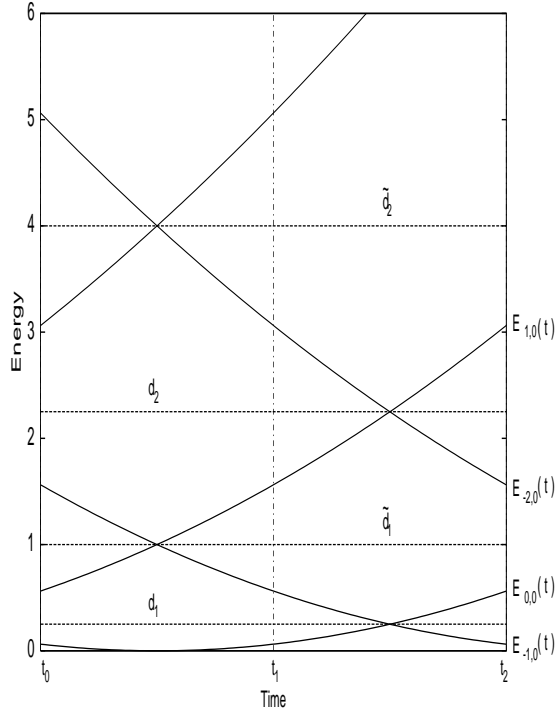}
\end{center}
\vskip -8cm
\caption{The first eigenvalues of the free Hamiltonian
$H_0(t)$ as a function of time and the vertical lines
$d_m$,${\tilde d}_m$ used in the text. }
\end{figure}

\begin{eqnarray}\label{2.24}
\sigma_{m,0}(t)&:=&E_{m-1,0}(t)\cup E_{-(m-1),0}(t), \qquad 
{\rm for}\,m\geq 1,\, \,\,  t\in I_0\\\label{2.25}
{\tilde \sigma}_{m,0}(t)&:=&E_{m-1,0}(t)\cup E_{-m,0}(t), \qquad  
{\rm for}\,m\geq 1,\, \,\, \,\,\, t\in I_1
\end{eqnarray}
so that $\sigma (H_0(t))=\cup_{m\geq 1}\sigma_{m,0}(t)$, 
$\sigma (H_0(t))=\cup_{m\geq 1}{\tilde \sigma }_{m,0}(t)$ on $I_0$ and $I_1$ respectively. An alternative 
way of labelling the eigenvalues of $H_0(t)$ is to count them in the increasing order 
\begin{equation}\label{2.26}
E_1^0(t)\leq E_2^0(t)
\leq E_3^0(t)...
\qquad t\in\Real .
\end{equation}
Of course the two labellings are related: for a fixed $\alpha $ 
and a given $t$ there exists an $n_{\alpha }(t)$ such that
\begin{equation}\label{labelling1}
E_{\alpha}^0(t)=E_{n_{\alpha }(t),0}(t)\quad {\rm and}\,\, 
P_{\alpha}^0(t)=P_{n_{\alpha }(t),0}.
\end{equation}
Let
$t\in [-\frac{1}{2},0)+\frac{N}{2}$ with $N=0,1,2,...$.
 Then the correspondence 
between $\alpha$ and $n_{\alpha }(t)$ is given as follows:  
\begin{eqnarray}\label{cor1}
n_{\alpha }(t)=\left\lbrace \begin{array}{ccc}
\frac{\alpha-N}{2},\qquad {\rm if}\,  N\, {\rm and}\, \alpha\, 
{\rm have\, the\, same\, parity}\\
\\   
-\frac{\alpha +N-1}{2},\qquad {\rm if }\, N\, {\rm and}\, \alpha\, 
{\rm have\, different\, parities.}
\end{array}\right.
\end{eqnarray}

With this labelling one has
\begin{eqnarray}
\sigma_{1,0}(t)&=&E_1^0(t);\quad \sigma_{m,0}(t)=E_{2m-2}^0\cup E_{2m-1}^0, 
\quad {\rm for}\,\,  t\in I_0\\\label{2.27} 
{\tilde\sigma }_{m,0}(t)&=&E_{2m-1}^0\cup E_{2m}^0, \quad {\rm for}\,\,  t\in I_1.
\end{eqnarray}
The spectral projections of $H_0(t)$ corresponding to $\sigma_{m,0}(t)$ 
and ${\tilde\sigma }_{m,0}(t)$ are denoted $Q_{m,0}$ and 
${\tilde Q}_{m,0}$:
\begin{eqnarray}
Q_{1,0}=P_{0,0};\quad Q_{m,0}&=&P_{m-1,0}+P_{-(m-1),0} \quad {\rm for}\,\,  t\in I_0\\
{\tilde Q}_{m,0}&=&P_{m-1,0}+P_{-m,0}\quad {\rm for}\,\,  t\in I_1.
\end{eqnarray}
By construction, for any $m=1,2...$ and $t\in I_0$, $ \sigma_{m,0}(t)$ 
are well separated . Similarly, for  $t\in I_1$  $ {\tilde\sigma}_{m,0}(t)$ 
are well separated:
\begin{eqnarray}\nonumber
\inf_{t\in I_0}{\rm dist }( \sigma_{m,0}(t), \sigma_{m+1,0}(t))&=& m-\frac{1}{2}\\
\label{2.28}
\inf_{t\in I_1}{\rm dist }( {\tilde\sigma}_{m,0}(t), {\tilde\sigma }_{m+1,0}(t))&=& m.
\end{eqnarray} 
Moreover, at the end of the interval $I_0$, the two eigenvalues 
composing $\sigma_{m,0}(t)$
(with the exception of $\sigma_{1,0}(t)$ which consists in just one eigenvalue)
  are well separated e.g.
\begin{equation}\label{2.29}
E_{2m-1}^0\left (-\frac{1}{4}\right )-E_{2m-2}^0\left (-\frac{1}{4}\right )= m-1 .
\end{equation}
The same is true for the two eigenvalues composing $ {\tilde\sigma}_{m,0}(t)$.
We come now to the spectral properties of $H(t)$.

Since $K(n+{1\over2},t)$ tends to zero  as $n$ tends to infinity, uniformly
with respect to $t$ the  analytic perturbation theory of type $B$, see [Kato,
chap VII], works here as long as $r>-{1\over2}$. More precisely if we label the eigenvalues of
$H(t)$ in increasing order
$E_{\alpha}(t),\alpha=1,...$ then $\lim_{\alpha\to\infty}(E_{\alpha}(t)-E_{\alpha}^0(t))=0$. In addition
for $t\neq 0,\frac{1}{2}$, when all the eigenvalues are nondegenerate the corresponding spectral 
projections
$P_{\alpha}(t)$ are close the unperturbed ones:
$\lim_{\alpha\to\infty}\|P_{\alpha}(t)-P_{\alpha}^0(t)\|=0$. 
In the following $\varphi_{\alpha}(t)$ denotes 
the eigenfunction of $H(t)$ corresponding to the eigenvalue $E_{\alpha}(t)$. 

Let 
\begin{equation}\label{2.30}
\Gamma _m=\{(m-\frac{1}{2})^2+iy|y\in\Real  \},\qquad
{\tilde\Gamma }_m=\{m^2+iy|y\in\Real  \} 
\qquad m=1,2,...   
\end{equation}
be vertical lines in the energy plane. 
In what follows, each time a quantity will be bounded from above
by a positive constant, this will be denoted by $C$, while for the lower bounds
we denote it $c$.

The main estimates, used many times along 
the paper are the following:  
 \vskip 0.2cm
\noindent {\sl {\bf Lemma 2.3:}
Let $A<\infty$.
Then for each $r>-\frac{1}{2}$ there exists a constant
$C_V$ independent of $m$ 
and a positive integer  
$N_r$ 
 such that for any $m>N_r$ the following estimate hold:
\begin{eqnarray}\nonumber
&&\sup_{t\in I_0}\left (\sup_{z\in\Gamma_m}\|K(z,t)\|
+\int_{\Gamma_m}\|R_0(z,t)\|\cdot\|K(z,t)\|dy  \right )\\\nonumber
&&+\sup_{t\in I_1}\left (\sup_{z\in{\tilde\Gamma }_m}\|K(z,t)\|
+\int_{{\tilde\Gamma}_m}\|R_0(z,t)\|\cdot\|K(z,t)\|dy  \right )\\\nonumber
&&+\max_{t=-\frac{1}{4},\frac{1}{4}}\left (\sup_{z\in\Gamma_m\cup {\tilde\Gamma }_m}\|K(z,t)\|
+\int_{\Gamma_m\cup {\tilde\Gamma}_m}\|R_0(z,t)\|\cdot\|K(z,t)\|dy  \right )\\\label{2.31}
&&\leq C_Vb(m),
\end{eqnarray}
where
\begin{equation}\label{2.32}
b(m):=\frac{
 \log ^2 \langle 4m-1 \rangle }{\langle m \rangle ^{1+\min\{0,2r\}}}.
\end{equation}
Also one has
\begin{equation}\label{2.32'}
\lim_{y\to\infty}\sup_{t\in I_0}\sup_{x\leq A}\|K(x+iy,t)\|=0.
\end{equation}
Let $m>1,$ $d_m:=(m-\frac{1}{2})^2$,
${\tilde d}_m:=m^2$ and $\gamma_m$, ${\tilde\gamma}_m$ be  closed finite contours
that intersect the real axis in and only in $d_m,d_{m-1}$ and ${\tilde d}_m,
 {\tilde d}_{m-1}$ (for example $\gamma_m$ can be a square of length $8(m-1)$).
Then there exist an absolute constant $C$ such that:
\begin{eqnarray}\label{aaa}
\sup_{t\in I_0}\sup_{z\in\gamma_m}\|K(z,t)\|&\leq& C\|K(d_m,0)\|,\\\label{bbb}
\sup_{t\in I_1}\sup_{z\in{\tilde\gamma}_m}\|K(z,t)\|&\leq& C\|K({\tilde d}_m,0)\|.
\end{eqnarray}

}

\noindent
Due to the above lemma we are able to give upper bounds of the type $(C_V\langle b(m) \rangle )^{N}$.
In the following we shall say that a quantity $A$ is of order 
$({\cal O}_r(\langle b(m) \rangle )^{N})$ if it obeys the estimate 
$\|A\|\leq (C_V\langle b(m) \rangle )^{N}$. The subscript
$r$ reminds us the dependence on $\|V\|_r$.   
We notice that the third line in (\ref{2.31}) follows at once from the estimates of the 
first and the second lines.  
As we have said, the proof of the Lemma is postponed to Appendix 7.1.
\\
\par
Now we are ready to spell the spectral properties of $H(t)$. We shall consider only $t\in I_0$;
the results and the proofs for $t\in I_1$, with the appropiate identifications, are the same. 
The first 
remark is that there exists an $m^*$ such that for any $m>m^*$ 
the norm of $K(z,t)$ is smaller
than $\frac{1}{2}$.
Secondly, for all $t\in I_0$ $d_m\in\rho(H(t))$ 
and  for all $t\in I_1$ ${\tilde d}_m\in\rho(H(t))$. Moreover, ${\tilde d}_m\in\rho(H(-\frac{1}{4}))$:

\vskip 0.2cm
\noindent {\sl {\bf Corollary 2.4:} Let $m>m^*$.
 Then 
\begin{equation}\label{2.33}   
\min \{\inf_{t\in I_0}{\rm dist}(d_m,\sigma(H(t))),
 \inf_{t\in I_1}{\rm dist}( {\tilde d}_m,\sigma(H(t))) \}\geq cm.
\end{equation}}
Let $Q_m(t)$ the spectral projection associated to 
  $\sigma_m(t)=(d_{m-1},d_m)\cap\sigma(H(t)) $.
Then one has the following:
\vskip 0.2cm  
\noindent {\sl {\bf Corollary 2.5:} For $m$ sufficiently large or $V$ small enough
\begin{equation}\label{2.34}
\sup_{t\in I_0}\|Q_m(t)-Q_{m,0}\|\leq C_Vb(m).
\end{equation}}
Let for $\alpha =2m-2,2m-1 $ (see (\ref{cor1}))
\begin{equation}\label{2.35}
P_{\alpha }^0\left (-\frac{1}{4} \right )=P_{n_{\alpha }(-\frac{1}{4}),0};
\quad \varphi_{\alpha }^0\left (-\frac{1}{4} \right )=\varphi_{n_{\alpha }
\left(-\frac{1}{4}\right ),0},
\end{equation}
and $P_{2m-2}\left (-\frac{1}{4} \right ),P_{2m-1}\left (-\frac{1}{4} \right )$
the spectral projections of $H\left (-\frac{1}{4} \right )$ corresponding to the 
intervals $(d_{m-1},{\tilde d}_{m-1})$ and  $({\tilde d}_{m-1},d_m)$ respectively.
Then the following estimate holds:  
\noindent
\vskip 0.2cm
\noindent {\sl {\bf Corollary 2.6:} Let $\alpha =2m-2,2m-1 $. Then for $m>m^*$
\begin{equation}\label{2.36}
\left \|P_{\alpha }\left (-\frac{1}{4} \right )
-P_{\alpha }^0\left (-\frac{1}{4} \right ) \right \|\leq C_Vb(m).
\end{equation}} 

\noindent
For sufficiently large $m$, by Sz-Nag\"{y} lemma (see Appendix 7.2), 
$P_{\alpha }\left (-\frac{1}{4} \right )$ and $P_{\alpha }^0\left (-\frac{1}{4} \right )$
 are unitarily equivalent so that  $H\left (-\frac{1}{4} \right )$ has nondegenerate
eigenvalues in $(d_{m-1},{\tilde d}_{m-1})$ and $({\tilde d}_{m-1},d_m)$ respectively.
 They are denoted by $E_{2m-2}\left (-\frac{1}{4} \right )$ and 
$E_{2m-1}\left (-\frac{1}{4} \right )$. Moreover, 
if $N_{\alpha }\left (-\frac{1}{4} \right )$ is the Sz-Nag\"{y} transformation
 corresponding to the pair $P_{\alpha }\left (-\frac{1}{4} \right )$,
 $P_{\alpha }^0\left (-\frac{1}{4} \right )$ then 
\begin{equation}\label{2.37}
\varphi_{\alpha }\left (-\frac{1}{4} \right )=N_{\alpha }\left (-\frac{1}{4} \right )
\varphi_{\alpha }^0\left (-\frac{1}{4} \right )
\end{equation}
are eigenvectors of  $H\left (-\frac{1}{4} \right )$ corresponding to 
$E_{\alpha }\left (-\frac{1}{4} \right )$. We say that an eigenbasis 
$\{\varphi_{\alpha }\left (-\frac{1}{4} \right ) \}_{\alpha =1}^{\infty }$ of
$H\left (-\frac{1}{4} \right )$ is a canonical one if for sufficiently 
large $\alpha$ (such that all the above construction works) 
$\varphi_{\alpha }\left (-\frac{1}{4} \right )$ is given by (\ref{2.37}). 
In the same way one constructs canonical bases for $t=\frac{1}{4}$. For 
$t_l:=-\frac{1}{4}+\frac{l}{2},l=2,3...$ the canonical bases are provided 
by 
(see (\ref{2.21}))
\begin{equation}\label{2.38} 
T\varphi_{\alpha }(t_l+1)=\varphi_{\alpha }(t_l).
\end{equation}
Before going to the proofs let us write the reduction to one period formula
(\ref{2.22}) in the canonical basis.

\vskip 0.2cm
\noindent {\sl {\bf Lemma 2.7 }
 Let $l$ be even, $t_0=-\frac{1}{4},t_l=t_0+\frac{l}{2}$. Then
for $\alpha$ sufficiently large 
\begin{equation}\label{2.39}
\langle \varphi _{\alpha }(t_l+1),U(t_l+1,t_l)\varphi _{\beta }(t_l)\rangle
=\langle \varphi _{\alpha}(t_2),U(t_2,t_0)
\varphi _{\beta }(t_0)\rangle .
\end{equation}}
Now we present the proofs for the lemmae and corollaries given above:
\vskip 0.2cm
\noindent
{\em Proof of Lemma 2.1:} We estimate the Hilbert-Schmidt norm of $K(-a^2,k,t)$

\begin{eqnarray*}
\|K(-a^2,k,t)\|^2&\leq& \|K(-a^2,k,t)\|_{HS}^2
\\
&=&\sum_{m,n\in\Int}\left |\frac{1}{\sqrt {(m+k+t)^2 +a^2}  }{\hat V}(m-n)
\frac{1}{\sqrt {(n+k+t)^2 +a^2}  } \right |^2  \\
&\leq&  \|V \|_r^2 \sum_{m,n\in\Int}\frac{1}{(m+k+t)^2 +a^2}
\langle m-n \rangle ^{-2r } \frac{1}{(n+k+t)^2 +a^2}
  \\ \nonumber
&\leq&  \|V \|_r^2\left (\frac{\sqrt{5}}{2}\right )^{\max\{-2r,0\}}\sum_{m\in\Int}
\frac{\langle m \rangle ^{\max \{-2r,0 \} }  }{ (m+k+t)^2 +a^2},
\end{eqnarray*}
where in the last line we used the inequality
\begin{equation}\label{ineq}
\bra m-n\ket^{-2r}\le \left (
\frac{\sqrt{5}}{2}\bra m\ket \bra n\ket\right )
^{\max\{-2r,0\}}.
\end{equation}
\par
\noindent
Then the above series is convergent for
  $r>-\frac{1}{2}$ and the lemma is proved
using the Lebesque dominated convergence theorem.\,\qed
\vskip 0.2cm

\noindent
{\em Proof of Corollary 2.4:} $R(d_m,t)$ can be written as
\begin{equation}\label{2.*1}
R(d_m,t)=|R_0(d_m,t)|^{\frac{1}{2}}(1+K(d_m,t))^{-1}R_0(d_m,t)^{\frac{1}{2}}.
\end{equation} 
Taking into account that ${\rm dist}(d_m,\sigma (H_0(t)))\geq cm $ one obtains from
(\ref{2.*1})
\begin{equation}
\|R(d_m,t)\|\leq 2\||R_0(d_m,t)|^{\frac{1}{2}}\|\cdot \|R_0(d_m,t)^{\frac{1}{2}}\|
=2\|R_0(d_m,t)\|\leq 2(Cm)^{-1},
\end{equation}
which implies $\inf_{t\in I_0}{\rm dist}(d_m,\sigma (H(t)))\geq cm$. The proof
of the inequality

 ${\rm dist}( {\tilde d}_m,\sigma(H(-\frac{1}{4})) \geq Cm/2$
is similar.\qed 
\vskip 0.2cm
\noindent
{\em Proof of Corollary 2.5:} By perturbation theory  
\begin{eqnarray*}
Q_m(t)-Q_{m,0}&=&\frac{i}{2\pi}\oint_{\gamma_m}(R(z,t)-R_0(z,t))dz\\  
&=&-\frac{i}{2\pi}\int_{\gamma_m}R_0(z,t)^{\frac{1}{2}}
K(z,t)(1+K(z,t))^{-1}|R_0(z,t)|^{\frac{1}{2}}dz.
\end{eqnarray*}
Taking the norms, using that $\oint_{\gamma_m}\|R_0(z,t)\|dz\leq C$ 
the estimate (\ref{2.34}) follows from (\ref{aaa})
and (\ref{2.31}).
\vskip 0.2cm
\noindent
{\em Proof of Corollary 2.6:} Similar to the proof of  Corollary 2.5.\qed
\vskip 0.2cm
\noindent
{\em Proof of Lemma 2.7:} Direct computation using  (\ref{2.22}),  (\ref{2.37}) 
and  (\ref{2.38}).\qed

\setcounter{equation}{0}
\section{The adiabatic theorem}
\par

As already said in the introduction our first task is to show 
that at high energies the most important "interband transitions" 
during a half Bloch period are the so-called Zener transitions, namely the ones 
 between neighbouring bands (e.g on $I_0$ between $ E_{2m-2}(t)$
and $ E_{2m-1}(t)$). This is nothing else but to claim that for $t,s\in I_0$ the 
subspace $Q_{m}(t){\cal H}$ is almost
invariant under the evolution $U(t,s)$.
\noindent
In this section we construct adiabatic evolutions $U^A_m(t,t_0)$ and 
${\tilde U}^A_m(t,t_1)$ on $I_0$  and $I_1$ respectively, satisfying
\begin{eqnarray}
U^A_m(t,t_0)Q_{m}(t_0){\cal H}=Q_{m}(t){\cal H},\qquad t\in I_0\\
{\tilde U}^A_m(t,t_1){\tilde Q}_{m}(t_1){\cal H}={\tilde Q}_{m}(t){\cal H},\qquad t\in I_1
\end{eqnarray}
and prove that they are close to $U(t,t_0)$ and $U(t,t_1)$ respectively.
We consider first the half Bloch period $I_0$ and at the end indicate the changes for $I_1$.
Let
\begin{eqnarray}
Q_{m}^+(t):=\sum_{j=m+1}^{\infty}Q_j(t),\qquad
Q_{m}^-(t):=\sum_{j=1}^{m-1}Q_j(t),\qquad t\in I_0,
\end{eqnarray}
and define $H^A_m(t)$ by
\begin{eqnarray}\nonumber
H^A_m(t)&:=&H(t)-X_{m}(t)\\
&=:& H(t)-iQ_{m}^+(t){\dot Q}_{m}^+(t)-iQ_{m}(t){\dot Q}_{m}(t)
-iQ_{m}^-(t){\dot Q}_{m}^-(t)\label{3.3}.
\end{eqnarray}
It turns out (see the proof of Lemma 3.4 below) that $X_{m}(t)$ are bounded hence $H^A_m(t)$
are self-adjoint on ${\cal D}(H(t))$. The adiabatic evolution $U^A_m(t,t_0)$
generated by $H^A_m(t)$ 
will be defined as the solution (in the weak sense, as in Lemma 2.2 ) of 
the equation
\begin{equation}\label{3.4}
i\frac{d}{dt}U^A_{m}(t,t_0)=H^A_{m}(t)U^A_{m}(t,t_0),
\qquad U^A_{m}(t_0,t_0)=1.
\end{equation}
As in the case of standard adiabatic theorem of quantum mechanics, 
$U^A_m(t,t_0)$ has the intertwining property
\begin{eqnarray}
 Q_{m}(t)=U^A_{m}(t,t_0)Q_{m}(t_0)U^A_{m}(t,t_0)^*,\qquad
 Q_{m}^{\pm}(t)=U^A_{m}(t,t_0)Q_{m}^{\pm}(t_0)U^A_{m}(t,t_0)^*,
 \end{eqnarray}
i.e the families of subspaces $Q_{m}(t){\cal H}$, $Q_{m}^{\pm}(t){\cal H}$ are 
invariant under $U^A_m(t,t_0)$. The proof is standard 
( Krein-Kato Lemma \cite{Nenciu1,KK,K})
For $t\in I_1$ one constructs in a 
similar way ${\tilde U}_{m}^A(t,t_1)$ by replacing $Q_{j}(t)$ by ${\tilde Q}_{j}(t)$. 
The main result of this section is to show that $U^A_m(t,t_0)$ and $U(t,t_0)$
are close:
\vskip 0.2cm 
\noindent {\sl {\bf Theorem 3.1} 
For any $ r>-\frac{1}{2}$ and $m$ sufficiently large or $\|V\|_r$ small enough
 it holds:
 \begin{eqnarray}\label{3.6}
 \sup_{t\in I_0}\|U(t,t_0)-U_{m}^A(t,t_0)\|
 \leq C_Vb(m)\langle m \rangle ^{-1},\\\label{3.7}
 \sup_{t\in I_1}\|U(t,t_1)-{\tilde U}_{m}^A(t,t_1)\|
  \leq C_Vb(m)\langle m \rangle ^{-1}.
\end{eqnarray}
where 
$C_V$ is a constant that depends on $V$ and $b(m)$ was introduced in Lemma 2.3.}
\\
\par
The basic steps of the proof are the standard ones but we have to check 
the relevant estimates. Define
\begin{equation}\label{3.8}
\Omega_{m}(t,t_0):=U^A_{m}(t,t_0)^*U(t,t_0),
\end{equation}
and notice that $\|\Omega_{m}(t,t_0)-1\|=\|U_{m}^A(t,t_0)-U(t,t_0)\|$,
so one is left with the estimation of $\|\Omega_{m}(t,t_0)-1\|$. 
Since a direct estimation of
\begin{equation}\label{3.9}
i(\Omega_{m}(t,t_0)-1)=\int_{t_0}^tdsU_{m}^A(s,t_0)^*
  X_{m}(s)U_{m}^A(s,t_0)\Omega_{m}(s,t_0)
\end{equation} 
would give (see Lemma 3.4) a bound of order 
${\cal O}_r(b(m))$, which is not accurate enough for the error estimates
in the long time behaviour 
we have to follow the proof of the standard adiabatic theorem and make 
"an integration by parts". More precisely we have: 
   \par
   \vskip 0.2cm
   \noindent {\sl {\bf Lemma 3.2:}
    $X_{m}(t)$  can be written in the form (we omit for simplicity the
  time-dependence):
    \begin{equation}\label{3.10}
    X_{m}=\left [H,Y_{m}\right ]
    \end{equation}
    with:
\begin{eqnarray}\nonumber
Y_{m}&=&\frac{1}{2\pi }\oint_{\gamma _{m}^-}dz
R(z)\left\lbrace Q_{m}^-{\dot Q}_{m}^--
Q_{m}^+{\dot Q}_{m}^+Q_{m}^-\right\rbrace R(z)\nonumber \\\label{3.11}
&+&\frac{1}{2\pi }\oint_{\gamma _{m}}dzR(z)
\left\lbrace Q_{m}{\dot Q}_{m}-Q_{m}^+
{\dot Q}_{m}^+Q_{m}
\right\rbrace R(z).
\end{eqnarray}}
Here $\gamma _{m}^-$ is a closed contour around $\sigma (Q_m^-H(t))$.
Now the integration by parts will give $\|\Omega_{m}(t,t_0)-1\|$ in terms of $Y_{m}$:

 \vskip 0.2cm
\noindent {\sl {\bf Lemma 3.3:} If $t$ belongs to $I_0$ the following estimate holds: 
\begin{equation}\label{3.12}
\|\Omega_{m}(t,t_0)-1)\|\leq \|Y_{m}(t)\|+\|Y_{m}(t_0)\|+
\int_{t_0}^tds \left (\|X_{m}(s)Y_{m}(s)\|+
\|{\dot Y}_{m}(s)\|\right ),
\end{equation}}
so we are left with verifying (\ref{3.10}-\ref{3.12}) and to estimate the norms 
in (\ref{3.12}). The basic estimation is contained in:

\vskip 0.2cm
\noindent {\sl {\bf Lemma 3.4:} For $m$ sufficiently large
\begin{equation}\label{3.13}
\sup_{t\in I_0}\left (\|{\dot Q}_{m}(t)\|+\|{\dot Q}_{m}^-(t)\|+
\|{\dot Q}_{m}^+(t)\| \right )
\leq C_Vb(m).
\end{equation}}
Notice that since $Q_{m}^+(t)=1-Q_{m}(t)-Q_{m}^-(t)$ we have only to estimate 
$\|{\dot Q}_{m}(t)\|$ and $\|{\dot Q}_{m}^-(t)\|$. Taking Lemma 3.4 for granted
we give now the proofs of Lemmae 3.2 and 3.3. 

\vskip 0.2cm
\noindent
{\em Proof of Lemma 3.2:}
Since by Lemma 3.4 $X_m(t)$ is bounded 
\begin{equation}\label{3.14}
\sup_{t\in I_0}\|X_m(t)\|\leq C_b(m)
\end{equation}
and in addition (remember that for an orthogonal projection valued function $E(t)$
 $E(t){\dot E}(t)E(t)=0$) $X_m(t)$ is off-diagonal with respect to the  decomposition
${\cal H}=Q_{m}^+(t){\cal H}\oplus Q_{m}{\cal H}\oplus Q_{m}^-{\cal H}$ one obtains 
$Y_m(t)$ as the solution of the commutator equation (\ref{3.10}) by using Theorem 9.3 in 
\cite{BhaRos}. Actually one can verify that (\ref{3.11}) solves \ref{3.10}) 
by straightforward computation.  
\qed
\vskip 0.2cm
\noindent
{\em Proof of Lemma 3.3:}
 Using the identity (we omit for simplicity the
 index $m$ and the time-dependence):
 \begin{eqnarray*}\nonumber
 i\frac{d}{dt}({U^A}^*YU^A)&=&{U^A}^*
 \left\lbrace i{\dot Y}-[H^A,Y]\right\rbrace {U^A}\\
 &=&{U^A}^*\left\lbrace i{\dot Y}-X+[X,Y]\right\rbrace {U^A}
 \end{eqnarray*}
   and replacing ${U^A}^*XU^A$ in (\ref{3.9}) one has:
\begin{eqnarray*}
i(\Omega-1)=\int_{t_0}^tds\left\lbrace -i\frac{d}{ds}({U^A}^*YU^A)+
{U^A}^*[X,Y]U^A+i{U^A}^*{\dot Y}U^A\right\rbrace \Omega \\
=\int_{t_0}^tds\left\lbrace -i\frac{d}{ds}({U^A}^*YU^A\Omega)
+i{U^A}^*YU^A{\dot \Omega}+{U^A}^*[X,Y]U^A\Omega+
i{U^A}^*{\dot Y}U^A\Omega\right\rbrace\\
=-i{U^A}^*YU^A\Omega\vert _{t_0}^t+\int_{t_0}^tds  {U^A}^*\left\lbrace
XY+i{\dot Y}\right\rbrace U^A\Omega
\end{eqnarray*}
which in norm gives the claimed result.\qed
\vskip 0.2cm
\noindent
{\em Proof of Lemma 3.4:} We give detailed calculations only for ${\dot Q}^{-}_m(t)$.
Computing $Q_{m}$ perturbatively, using the identity
${\dot R}(z,t)=-R(z,t){\dot H}(t)R(z,t)$ 
 one has
 \begin{eqnarray}\nonumber
 {\dot Q}_{m}^-(t)&=&\frac{d}{dt}\left\lbrace
  Q_{m,0}^--\frac{i}{2\pi }\oint_{\gamma _{m}^-}dz
     R_0(z)VR(z)
      \right\rbrace\\ \nonumber
       &=&\frac{i}{2\pi }\oint_{\gamma _{m}^-}dz
        \left\lbrace R_0(z){\dot H}_0R_0(z)
         VR(z)+R_0(z)VR(z)
          {\dot H}R(z) \right\rbrace ,
           \end{eqnarray}
where we omited the explicit $t$-dependence.
 To go further with the estimates we deform first $\gamma _{m}^-$ to
$\gamma _{m}^-(L):=\gamma_{m-1}(L)\cup\gamma '_{m-1}(L)\cup\gamma ''_{m-1}(L)
\cup\gamma_0(L)$ with:
\begin{eqnarray}
\gamma_{m-1}(L)&:=&\{d_{m-1}+iy|y\in [-L,L] \}\\
\gamma '_{m-1}(L)&:=&\{x+iL|x\in [-L,d_{m-1}]\}\\
\gamma ''_{m-1}(L)&:=&\{x-iL|x\in [-L,d_{m-1}]\}\\
\gamma_0(L)&:=&\{-L+iy|y\in [-L,L] \},
\end{eqnarray}
and then use the estimate (\ref{2.32'})
to show that the only integral that survives from $\gamma _{m}^-$
in the limit $L\to\infty $ is the one on $\Gamma_{m-1}$ (notice that
$\lim_{L\to\infty }\gamma_{m-1}(L)=\Gamma_{m-1}$).
 Using the polar decomposition
of $R_0(z)$ and the resolvent equation for $R(z)$ 
it follows that
\begin{eqnarray}\nonumber
\|{\dot Q}_{m}^-(t)\|&\leq&\frac{1}{2\pi}
 \oint_{
\Gamma _{m-1}}|dz|\|R_0(z)\|\|K(z)\|\cdot
\||R_0(z)|^{\frac{1}{2}}{\dot H}_0R_0(z)^{\frac{1}{2}}\|\cdot 
\|(1+K(z))^{-1}\|\\\nonumber
&+&\frac{1}{2\pi} \oint_{
\Gamma _{m-1}}|dz|\|R_0(z)\|\|K(z)\|
\cdot\||R_0(z)|^{\frac{1}{2}}{\dot H}_0R_0(z)^{\frac{1}{2}}\|
\cdot\|(1+K(z))^{-1}\|^2.
\end{eqnarray}
We note that the quantity  $|R_0(z,t)^{\frac{1}{2}}|{\dot H}_0(t)R_0(z,t)
^{\frac{1}{2}}$ is uniformly bounded on $\Gamma_{m-1}$ (namely this quantity is
 bounded in norm by a constant). Indeed:
\begin{eqnarray*}
\sup_{t\in I_0}\sup_{z\in\Gamma_{m-1}}
\|R_0(z,t)^{\frac{1}{2}}{\dot H}_0(t)|R_0(z,t)|^{\frac{1}{2}} \|
&=&\sup_{t\in I_0}\sup_{y\in\Real }\sup_{m'\in \Int}
\left |\frac{2(m'+t)}{(m'+t)^2-d_{m-1}-iy}\right|\\
&\leq & C.
\end{eqnarray*}
Then using Lemma 2.3 and the inequality $\|(1+K)\|^{-1}\leq (1-\|K\|)^{-1}$ one obtains 
\begin{equation}
\sup_{t\in I_0}\|{\dot Q}_{m}^-(t)\|\leq C_Vb(m-1). 
\end{equation}
Since $b(m-1)\leq Cb(m)$ we proved the claimed result on ${\dot Q}_{m}^-(t)$. The estimate of ${\dot Q}_{m}(t)$
goes along the same line, by deforming the closed contour around $\sigma (Q_{m}(t)H)$,
$\gamma_m$, to $\Gamma_{m-1}\cup\Gamma_{m} $. The lemma is now proved\qed

\vskip 0.2cm
\noindent
We turn now to the estimations of the norms in the r.h.s of (\ref{3.12}). For
$X_m(t)$ Lemma 3.4 gives:  
\begin{equation}\label{ad1}
\sup_{t\in I_0}\|X_m(t)\|\leq C_Vb(m),
\end{equation}
while for $Y_m(t)$ we write
\begin{eqnarray}\nonumber
\|Y_{m}(t)\|
&\leq &\frac{1}{2\pi } \left(\|{\dot Q}_{m}^-(t)\|
+\|{\dot Q}_{m}^+(t)\|\right )
\int_{\Gamma _{m-1}}dz\|R(z,t)\|^2 \\
&+& \frac{1}{2\pi }\left(\|{\dot Q}_{m}(t)\|+\|{\dot Q}_{m}^+(t)\|\right )
\int_{\Gamma _{m-1}\cup\Gamma _{m}}dz\|R(z,t)\|^2.
\end{eqnarray}
Now for example (use again the estimate of $K(z,t)$ on $\Gamma _{m-1}\cup\Gamma _{m}$)
\begin{eqnarray*}
&&\sup_{t\in I_0}\int_{\Gamma _{m-1}\cup\Gamma _{m}}|dz|
\|R(z,t)\|^2\leq C \sup_{t\in I_0}\int_{\Gamma _{m-1}\cup\Gamma _{m}}
|dz|\|R_0(z,t)\|^2\\
&=&\sup_{t\in I_0}\sup_{m'\in\Int }\int_{-\infty}^{\infty}
\left\lbrace\frac{dy}{[E_{m',0}(t)-d_{m-1}]^2+y^2}
+ \frac{dy}{[E_{m',0}(t)-d_{m}]^2+y^2}  \right\rbrace 
\leq C\langle m \rangle ^{-1},
\end{eqnarray*}
which together with Lemma 3.4 gives
\begin{equation}\label{ad2}
\sup_{t\in I_0}\|Y_m(t)\|\leq C_Vb(m)\langle m\rangle ^{-1}.
\end{equation}
Obviously, $Y_{m}(t_0)$ satisfies the same estimate.
To estimate $\|{\dot Y}_{m}(t)\|$
we observe first that it involves estimates on
${\ddot Q}_{m}^-(t),{\ddot Q}_{m}^+(t) $
and ${\ddot Q}_{m}(t)$. Indeed, a long but straightforward 
calculation gives (we omit the index $m$ in $Y_m$):
\begin{eqnarray*}
{\dot Y}(t)&=&
-\frac{1}{2\pi i}
\int_{\Gamma _{m-1}}dz\left\lbrace {\dot R}(z,t)
\left (Q^-(t){\dot Q}^-(t)-
Q^+(t){\dot Q}^+Q^-\right ) R(z,t)\right . \nonumber\\
&+&R(z,t)(({\dot Q}^-(t))^2+Q^-(t)
{\ddot Q}^-(t)-({\dot Q}^+(t))^2Q^-(t)\nonumber\\
&-&Q^+(t){\ddot Q}^+(t)Q^-(t)-Q^+(t)
{\dot Q}^+(t){\dot Q}^-(t))R(z,t)
\nonumber\\
&+&\left .   R(z,t)\left ( Q^-(t){\dot Q}^-(t)-
Q^+(t){\dot Q}^+Q^-(t)\right ) {\dot R}(z,t)\right\rbrace \nonumber\\
&-&\frac{1}{2\pi i}\int_{\Gamma _{m-1}\cup \Gamma _{m} }dz\left\lbrace
{\dot R}(z,t)\left ( Q(t){\dot Q}(t)-Q^+(t){\dot Q}^+(t)Q(t)\right )R(z,t)\right . \nonumber\\
&+&R(z,t) ({\dot Q}(t)^2+Q(t){\ddot Q}(t)-
{\dot Q}^+(t)^2Q(t)\nonumber\\
&-&Q^+(t){\ddot Q}^+(t)Q(t)
-Q^+(t){\dot Q}^+(t){\dot Q}(t) )R(z,t)\nonumber\\
&+&\left .R(z,t)\left (Q(t){\dot Q}(t)-Q^+(t){\dot Q}^+(t)
Q(t)\right ){\dot R}(z,t)\right\rbrace .
\end{eqnarray*}
In what concerns ${\ddot Q}_{m}^-(t)$, one has only to perform all the derivatives
involved
and then to estimate the terms
\begin{eqnarray*}
{\ddot Q}_{m}^-(t)
 &=&\frac {1}{2\pi i}
 \int_{\Gamma _{m-1}}dz \left\lbrace
 R_0(z,t){\dot H}_0(t)R_0(z,t){\dot H}_0(t)R_0(z,t)VR(z,t) \right .\\
 &-&R_0(z,t){\ddot H}_0(t)R_0(z,t)VR(z,t)\\
 &+&R_0(z,t){\dot H}_0(t)R_0(z,t){\dot H}_0(t)R_0(z,t)VR(z,t)\\
 &+&R_0(z,t){\dot H}_0(t)R_0(z,t)VR(z,t){\dot H}_0R(z,t)\\
 &+&R_0(z,t){\dot H}_0(t)R_0(z,t)VR(z,t){\dot H}_0(t)R(z,t)\\
 &+&R_0(z,t)VR(z,t){\dot H}_0(t)R(z,t){\dot H}_0(t)R(z,t)\\
 &-&R_0(z,t)VR(z,t){\ddot H}_0(t)R(z,t)\\
 &+&\left .R_0(z,t)VR(z,t){\dot H}_0(t)R(z,t){\dot H}_0(t)R(z,t)\right\rbrace .
 \end{eqnarray*}
  We claim that the estimates for these terms involve only powers of
  $\left (1+ K(z,t)\right ) ^{-1}$, $R_0(z,t)^{\frac{1}{2}}
  {\dot H}_0(t)|R_0(z,t)|^{\frac{1}{2}}$  and the integral
  of $\|R_0(z,t)^{\frac{1}{2}}\|^2\cdot\|K(z,t)\| $ on $\Gamma_{m-1}$
   which was already estimated in Lemma 2.3. The result is 
\begin{equation}
\sup_{t\in I_0}\|{\ddot Q}_{m}^-(t)\|\leq C_Vb(m).
\end{equation}
The same estimate is satisfied by ${\ddot Q}_{m}(t)$ and then  
\begin{equation}\label{ad3}
\sup_{t\in I_0}\|{\dot Y}(t)\|\leq C_Vb(m)\langle m \rangle ^{-1}.
\end{equation}
 The last term to be considered
  is $X_{m}(t)Y_{m}(t)$ which gives a better estimate
  \begin{equation}\label{ad4}
  \sup_{t\in I_0}\|X_{m}(t)Y_{m}(t)\|
  \leq C_V^2b(m)^2\langle m\rangle ^{-1}.
  \end{equation}
Collecting  (\ref{3.12}),(\ref{ad1}),(\ref{ad2}),(\ref{ad3}) and (\ref{ad4}) one obtains the first estimate 
from Theorem 3.1.
 For ${\tilde U}_{m}^A(t,t_0)$ the computations and estimations
 are similar as above, the only difference being that the contour of integration
 ${\tilde \Gamma }_{m}$ is slightly
 different, i.e it is shifted upwards with $\frac{1}{2}$ with respect to
 $\Gamma _{m}$. As a consequence all the estimations will
 be improved with $\frac{1}{2}$,
 in the sense that $b(m)\to b(m+\frac{1}{2})$. Thus there is no loss
 if we write
 \begin{equation}
 \sup_{t\in I_1}\|U(t,t_0)-{\tilde U}_{m}^A(t,t_0)\|
 \leq C_Vb(m)\langle m\rangle ^{-1}
 \end{equation}
The proof of Thm 3.1 is done. \qed 
\\
\noindent

\setcounter{equation}{0}
\section{ The adiabatic evolution. The effective Hamiltonian }
\par

As shown in Theorem 3.1 the true evolution over the half-periods $I_0$ and $I_1$
is well approximated, in the limit of large $m$, by the adiabatic evolutions
$U^A_m$ and ${\tilde U}^A_m$. The next step is to compute, up to small controlable errors, 
$U^A_m$ when restricted to $Q_m(t){\cal H}$ and 
${\tilde U}^A_m$  when restricted to ${\tilde Q}_m(t){\cal H}$. 
In what follows we consider explicitly 
only $U^A_m$ over $I_0$ and give the similar results for ${\tilde U}^A_m$ over $I_1$.  

Before entering the computation of $U^A_m$ let us show that, when restricted
to $Q_m(t){\cal H}$, $U^A_m$ is nothing but the simplified dynamics in \cite{Ao} 
obtained by keeping only the couplings between almost touching bands. Indeed, consider
an initial wave function $\psi_m(s)\in Q_m(s){\cal H}$, $s\in I_0$. Then, as far as 
$t\in I_0$, the adiabatic vector 
$\psi^A_{m}(t):=U^A_{m}(t,s)\psi_m (s)\in Q_m(t){\cal H}$ and  
it can be written as
\begin{equation}\label{4.1}
\psi^A_{m}(t)=\sum_{j=1}^{2}c_{j}(t)\varphi_{\alpha_j}(t),\qquad \alpha_1=2m-2, \alpha_2=2m-1,
\end{equation} 
where $\varphi_{\alpha_j}(t)$ are eigenfunctions of $H(t)$ corresponding 
to the eigenvalues from $\sigma (H(t))$ (one can assume that for $t=t_0$ or $t=t_1$, 
$\varphi_{\alpha_j}(t)$
coincide with the canonical eigenfunctions constructed in Section 2). 
Plugging the decomposition (\ref{4.1}) into the Schr\"{o}dinger equation for 
$\psi^A_{m}(t)$ (see Section 3)

\begin{equation}\label{4.2}
i\frac{d}{dt}\psi^{A}_{m}(t)=H^A_{m}(t)\psi^{A}_{m}(t)=(H(t)-X_m(t))\psi^{A}_{m}(t),
\end{equation} 
taking the scalar product with $\varphi_{\alpha_l}(t)$ and using the fact that 
$X_m(t)$ is off-diagonal i.e $Q_{m}(t)X_m(t)Q_{m}(t)=0$, one obtains:
\begin{equation}\label{4.3}
i\frac{d}{dt}c_l(t)=c_l(t)E_{\alpha_l}(t)+\sum_{j=1}^{2}\chi_{l,j}(t)c_j(t), 
\end{equation}
where 
\begin{equation}\label{4.4}
\chi_{l,j}(t):=-i\langle\varphi_{\alpha_l}(t),\frac{d}{dt}\varphi_{\alpha_j}(t)\rangle
\end{equation}
which is nothing but the equation given in Ao's paper.
\\
To compute $U^A_m$ amounts then to solve the $2\times 2$ system (\ref{4.3}) and this 
problem has been considered many times (see \cite{Ao,Br,LH} and references therein)
in the physical literature. 
The trouble with  (\ref{4.3}) is that $\chi_{l,j}(t)$ are not easy to compute
up to a controled error in the limit $m\to\infty $ since due to the existence 
of quasi-crossings one has to deal with almost degenerate perturbation theory.
To our knowledge the earlier papers take for $\chi_{l,j}(t)$ a low order 
perturbation theory formula, but never  controlled the rest. We avoid this difficulty
by using the reduction theory which is the standard tool in analytic perturbation 
theory \cite{K} and was as well extended to other contexts, in particular for 
adiabatic perturbation 
theory (see e.g \cite{Nenciu1,MN} and references therein). 
More precisely, for $m$ sufficiently large (see Corollary 2.2) $\|Q_m(t)-Q_{m,0}\|<1 $
and then (see Appendix 7.2) one can write the Sz-Nag\"y transformation matrix
corresponding to the pair $Q_m(t),Q_{m,0}$: 
\begin{equation}\label{4.6}
Q_m(t)=W_m(t)Q_{m,0}W_m(t)^*.
\end{equation}
Then if one defines for $t,s\in I_0$:
\begin{equation}\label{4.7}
{\cal U}_m(t,s):=W_m(t)^*U^A_m(t,s)W_m(s),
\end{equation}
it can be checked by straightforward calculation 
that 
\begin{equation}\label{4.8}
i\frac{d{\cal U}_m(t,s)}{dt}=\left ( i{\dot W}_{m}(t)^*W_{m}(t)
+W_{m}(t)^*H^A_{m}(t)W_{m}(t)\right ){\cal U}_m(t,s)
\end{equation} 
and that
\begin{equation}\label{4.9}
[Q_{m,0},{\cal U}_m(t,s)]=0.
\end{equation}
As a consequence if by definition 
\begin{equation}\label{4.10}
{\cal U}_{{\rm eff},m}(t,s):=Q_{m,0}{\cal U}_m(t,s)Q_{m,0}
\end{equation}
then ${\cal U}_{{\rm eff},m}(t,s)$ satisfies 
(as operators in $Q_{m,0}{\cal H}$)  
the equation of motion:
\begin{eqnarray}\nonumber
i\frac {d{\cal U}_{{\rm eff},m}(t,s)}{dt}&=&
Q_{m,0}\left(i{\dot W}_{m}(t)^*W_{m}(t)+W_{m}(t)^*H(t)W_{m}(t)\right )
Q_{m,0}{\cal U}_{{\rm eff},m}(t,s)\\\label{4.11}
&=&H_{{\rm eff},m}(t){\cal U}_{{\rm eff},m}(t,s).
\end{eqnarray}
Going backwards, once ${\cal U}_{{\rm eff},m}(t,s)$ is known one can 
recover $U^A_m(t,s)$ when restricted to $Q_m(t_0){\cal H}$: 
\begin{equation}\label{4.12}
U^A_m(t,t_0)Q_m(t_0)=W_{m}(t){\cal U}_{{\rm eff},m}(t,t_0)W_{m}(t_0)^*Q_m(t_0).
\end{equation}
The point of the reduction theory outlined above is that we can compute 
$H_{{\rm eff},m}(t)$ up to a controlled error:
\par
\vskip 0.2cm
\noindent {\sl {\bf Theorem 4.1:} 
For sufficiently large $m$ 
\begin{eqnarray}
H_{{\rm eff},m}(t)&=&\nonumber
Q_{m,0}\left (H_0(t)+V\right )Q_{m,0}\\\label{4.13}
&+&\frac{1}{2}Q_{m,0}\left ({\hat E}_{m,1}(t)V+V{\hat E}_{m,1}(t)\right )Q_{m,0}+
\Delta H_{{\rm eff},m}(t),
\end{eqnarray}
where
\begin{equation}\label{4.14}
{\hat E}_{m,1}(t):=\frac{1}{2\pi i}\oint_{\gamma _{m}}
dzR_0(z,t)VR_0(z,t)
\end{equation}
and
\begin{equation}\label{4.15}
\sup_{t\in I_0}\| \Delta H_{{\rm eff},m}(t)\|
\leq C_V\langle m \rangle b(m)^3.
\end{equation}}
We prove the theorem in few lemmae, each lemma giving an estimate for 
different terms
that appear in $H_{{\rm eff},m}(t)$. In the proofs we shall use several 
results of the perturbation theory. As well known, $Q_{m}(t)$ has 
the following expansion  
\begin{equation}
Q_{m}(t)=\sum_{j=0}^{N}{\hat E}_{m,j}(t)+E_{m,N+1}(t),
\end{equation}
with 
\begin{eqnarray}\label{4.21}
{\hat E}_{m,j}(t):=\frac{(-1)^ji}{2\pi}\oint_{\gamma_{m}}dzR_0(z,t)
(VR_0(z,t))^j,\\\label{4.22}
E_{m,j}(t):=\frac{(-1)^ji}{2\pi}\oint_{\gamma_{m}}dz(R_0(z,t)V)^j
R(z,t).
\end{eqnarray}
Using the identity $Q_{m}(t)=(Q_{m}(t))^2$ one can easily check the 
following relations
\begin{eqnarray}\label{i1}
Q_{m,0}{\hat E}_{m,1}(t)Q_{m,0}&=&0,\\\label{i2}
Q_{m,0}{\hat E}_{m,2}(t)Q_{m,0}&=&-Q_{m,0}{\hat E}_{m,1}^2(t)Q_{m,0},
\\\label{i3}
[H_0(t),{\hat E}_{m,1}(t)]&=&-[V,Q_{m,0}].
\end{eqnarray}
Moreover, $E_{m,j}(t)$ and its derivatives obey the 
estimates
\begin{eqnarray}\label{E_j}
\sup_{t\in I_0}\| E_{m,j}(t)\|&\leq &C_V \left (b(m)\right )^j,\\
\label{dotE_j}
\sup_{t\in I_0}\|{\dot E}_{m,j}(t)\|&\leq &C_V (j+1)\left (b(m)\right )^j.
\end{eqnarray}
To see (\ref{E_j}) we deform the contour $\gamma_m$ into 
$\Gamma _{m-1}\cup\Gamma _{m}$ and write
\begin{eqnarray*}
\sup_{t\in I_0}\| E_{m,j}(t)\|&=&\frac{1}{2\pi}\sup_{t\in I_0}
\|\int_{\Gamma _{m-1}\cup\Gamma _{m}}dz
|R_0(z,t)|^{\frac{1}{2}}(K(z,t))^j (1+K(z,t))^{-1}R_0(z,t)^{\frac{1}{2}}\|\\
&\leq &\frac{2}{2\pi}\sup_{t\in I_0}\sup_{z\in\Gamma _{m-1}\cup\Gamma _{m}}
\|(K(z,t))^{j-1}\|\int_{\Gamma _{m-1}\cup\Gamma _{m}}|dz|
\|R_0(z,t)^{\frac{1}{2}}\|^2\|K(z,t)\| \\
&\leq & C_V \left (b(m)\right )^j,
\end{eqnarray*}
where in the last line we used Lemma 2.3.
The estimate (\ref{dotE_j}) is obtained in the same way. 
We remark without giving
details that ${\hat E}_{m,j}$ and ${\dot {\hat E}}_{m,j}$ verify the same estimates
as $E_{m,j}$ and ${\dot E}_{m,j}$. 
Now we list the estimates for the various term appearing in $H_{{\rm eff},m}$: 

\vskip 0.2cm
\noindent {\sl {\bf Lemma 4.2:} The first term in the effective 
Hamiltonian obeys the following estimate:
\begin{equation}
\sup_{t\in I_0}\|Q_{m,0}{\dot W}_{m}(t)^*W_{m}(t)Q_{m,0}\|
\leq C_V\langle m\rangle b(m)^3 
\end{equation}}

\vskip 0.2cm
\noindent {\sl {\bf Lemma 4.3}: The second term is estimated as follows
\begin{eqnarray}\nonumber
Q_0W_m^*(t)H_0(t)W_m(t)Q_0
&=&Q_0H_0(t)Q_0-\frac{1}{2}
\left\lbrace Q_0V{\hat E}_{m,1}(t)Q_0+Q_0{\hat E}_{m,1}(t)VQ_0 \right \rbrace 
\\\label{lemma4.4}
&+&{\cal O}_r\left (\langle m\rangle b(m)^3\right )
\end{eqnarray}}

\vskip 0.2cm
\noindent {\sl {\bf Lemma 4.4}: The last term gives
\begin{eqnarray}\nonumber
Q_0W(t)_m^*VW_m(t)Q_0&=&Q_0VQ_0+Q_0(V{\hat E}_{m,1}(t)
+{\hat E}_{m,1}(t)V)Q_0
+{\cal O}_r\left (\langle m\rangle \left (b(m)\right)^3\right )\\
\end{eqnarray}}
Combining now Lemmae 4.2, 4.3 and 4.4 one obtains the explicit form
of the effective Hamiltonian as given in Theorem 4.1.
When proving the lemmae we skip for simplicity the index $m$ and the
time dependence of $W_m(t),E_{m,1}(t),Q_m(t)$ and other related quantities.
\vskip 0.2cm
\noindent
{\em Proof of Lemma 4.2:}
We shall use the fact that $E_1$, $\hat E_1$ and $E_2$ are symmetric hence in
 particular
$\partial_t(1-E_1^2)^{-{1\over2}}$ is also  symmetric. As we have said it is 
easy to see that   
$$
E_1=\OO_r(b(m)),\quad \hat E_1=\OO_r(b(m))\quad
E_2=\OO_r(b^2(m)).
$$
Denoting $L:=(1-E_1^2)^{-{1\over2}}=:1+M:=1+E_1^2F$ one has:
$$
WQ_0=LQQ_0,\quad Q_0\dot W^*=Q_0\dot QL+Q_0 Q\dot L
$$
and
$$
M=\OO_r(b^2(m)).
$$
Actually by writing $L^2=1+{\tilde N}$ we obtain  
\begin{eqnarray*}
Q_0\dot W^* WQ_0&=&{1\over2}Q_0\left(\dot W^* W-W^* \dot
W\right)Q_0\\
&=&{1\over2}Q_0\left(\dot QL^2Q-QL^2\dot Q\right)Q_0+{1\over2}Q_0Q[\dot
L,L]QQ_0\\
&=&{1\over2}Q_0[\dot Q,Q]Q_0
+{1\over2}Q_0\left(\dot Q{\tilde N}Q-Q{\tilde N}\dot Q\right)Q_0+{1\over2}Q_0Q[\dot
{\tilde N},{\tilde N}]QQ_0\\
&=&{1\over2}Q_0[\dot Q,Q]Q_0+\OO_r(b^3(m))
\end{eqnarray*}
where we used that ${\tilde N}$ and $\dot {\tilde N}$  are of $\OO_r(b^2(m))$ and $\dot
Q=\OO_r(b(m))$.
Now one should use the expansion $Q=Q_0+\hat E_1+E_2$ and the 
property $Q_0{\hat E_1} Q_0=0$
\begin{eqnarray*}
Q_0[\dot Q,Q]Q_0&=&Q_0[\dot {\hat E}_1+\dot E_2,Q_0+ {\hat E_1}+E_2]Q_0\\
&=&Q_0[\dot {\hat E}_1, {\hat E_1}]Q_0+\OO_r(b^3(m)).
\end{eqnarray*}
The last thing to be done is to show that $Q_0[\dot{\hat E}_1, {\hat E_1}]Q_0=0$
This follows by writing explicitely ${\hat E}_1$ (that is, by using residue theorem)
and by direct calculation.   
\qed   
\par

\vskip 0.2cm 
\noindent
{\em Proof of  Lemma 4.3: } 
The idea behind the proof is to write
\begin{equation}
Q_0W^*H_0WQ_0=Q_0W^*{\hat H}_0WQ_0+(m-1)^2Q_0
\end{equation} 
with ${\hat H}_0:=H_0-(m-1)^2Q_0$ and to estimate the first term. 
  As
$W=(1+E_1^2F)(QQ_0+(1-Q)(1-Q_0))$ then
\begin{equation}\label{WQ_0}
WQ_0=QQ_0+E_1^2FQQ_0
\end{equation}
from where
\begin{eqnarray*} 
Q_0W^*{\hat H}_0WQ_0 &=&Q_0Q(1+FE_1^2){\hat H}_0(1+E_1^2F)QQ_0\\
&=&Q_0Q{\hat H}_0QQ_0+Q_0Q({\hat H}_0E_1^2F+h.c)
+Q_0QFE_1^2{\hat H}_0E_1^2FQQ_0
\end{eqnarray*} 
We start by estimating the last term
\begin{equation}
\sup_{t\in I_0}\|FE_1^2{\hat H}_0E_1^2F\|\leq ct.\|E_1^2\|\cdot
\|E_1{\hat H}_0E_1\|,
\end{equation}
which requires an estimate on $E_1{\hat H}_0E_1$. To obtain it we shall
use 
\begin{equation}
\sup_{t\in I_0}\sup_{z,z'\in \Gamma _{m-1}\cup \Gamma _{m} }\|R_0(z',t)^{\frac{1}{2}}
{\hat H}_0R_0(z,t)^{\frac{1}{2}} \|\leq C.
\end{equation}
It follows from this that 
\begin{eqnarray*}\nonumber
&&\sup_{t\in I_0}\|E_1{\hat H}_0E_1\|\\
&=&\sup_{t\in I_0}\|\frac{1}{2\pi ^2}\oint _{\gamma _{m}}
\oint _{\gamma _{m}}dzdz'
R(z,t)VR_0(z,t){\hat H}_0R_0(z',t)VR(z',t)\|\\ \nonumber
&=&\sup_{t\in I_0}\|\frac{1}{2\pi ^2}\oint _{\gamma _{m}}
\oint _{\gamma _{m}}dzdz'
R_0(z,t)^{\frac{1}{2}}\left (1+K(z,t)\right )^{-1}K(z,t)
|R_0(z,t)|^{\frac{1}{2}}\cdot \\ \nonumber
&\cdot& {\hat H}_0|R_0(z',t)|^{\frac{1}{2}}
K(z',t)\left (1+K(z',t)\right )^{-1}|R_0(z',t)|^{\frac{1}{2}}\\
&\leq&C\oint _{\gamma _{m}}|dz|\|R_0(z,t)^{\frac{1}{2}}\|\|K(z,t)\|
\cdot \oint _{\gamma _{m}}|dz'| 
\|R_0(z',t)^{\frac{1}{2}}\|\|K(z',t)\| \\
&\leq& C_V\langle m\rangle b(m)^2.
\end{eqnarray*}
Then the term $FE_1^2{\hat H}_0E_1^2F$ turns out to be of order 
${\cal O}_r(\langle m\rangle b(m)^4)$ and can be disregarded.  
We continue the estimations with $Q_0QFE_1^2{\hat H}_0QQ_0$ using 
the perturbation theory up to $E_1$ (i.e $Q=Q_0+E_1$):
\begin{eqnarray*}
Q_0QFE_1^2{\hat H}_0QQ_0&=&Q_0FE_1^2{\hat H}_0Q_0
+Q_0FE_1^2{\hat H}_0E_1Q_0\\
&+&Q_0E_1FE_1^2{\hat H}_0E_1Q_0
+Q_0E_1FE_1^2{\hat H}_0Q_0.
\end{eqnarray*}
The second and the third terms in the above equation are simply 
estimated using the bounds on $E_1$ and $E_1{\hat H}_0E_1$.
For the fourth  term we have to estimate instead $Q_0{\hat H}_0E_1$. 
Following the same steps as in the estimation of $E_1{\hat H}_0E_1$
we arrive at
\begin{eqnarray}\nonumber
\sup_{t\in I_0}\|Q_0{\hat H}_0E_1(t)\|&=&
\sup_{t\in I_0}
\|\frac{1}{2\pi ^2}\oint _{\gamma _{m}}
\oint _{\gamma _{m}}dzdz'
R_0(z,t){\hat H}_0(t)R_0(z',t)V(z',t))R(z',t)\|\\\label{QHE_1}
&\leq& C_V\langle m\rangle b(m).
\end{eqnarray}
 In the similar way one can prove a more general result 
that will be used below, namely
\begin{equation}\label{QHE_j}
\sup_{t\in I_0}\|Q_0{\hat H}_0E_j\|
\leq C_V\langle m\rangle \left (b(m)\right )^j.
\end{equation}
It follows then from (\ref{QHE_1}) that 
\begin{eqnarray}\nonumber
\sup_{t\in I_0}\|Q_0E_1FE_1^2{\hat H}_0Q_0\|
&\leq & C\|E_1\|^2\|E_1{\hat H}_0Q_0 \|
\leq C_V\langle m\rangle \left (b(m)\right )^3.
\end{eqnarray}      
Hence a preliminar result is
\begin{eqnarray}\nonumber
Q_0W^*{\hat H}_0WQ_0 &=&Q_0Q{\hat H}_0QQ_0+Q_0Q(FE_1^2{\hat H}_0+
{\hat H}_0E_1^2F)QQ_0\\
&+&{\cal O}_r\left (\langle m\rangle 
\left (b(m)\right )^3 \right ). 
\end{eqnarray}    
To go further we shall use that $F=\frac{1}{2}+E_1^2G$ 
($G$ and its derivatives being again uniformely bounded). Then
\begin{eqnarray}\nonumber
Q_0QFE_1^2{\hat H}_0QQ_0=\frac{1}{2}Q_0QE_1^2{\hat H}_0QQ_0
+Q_0QE_1^2GE_1^2{\hat H}_0QQ_0
\end{eqnarray}
and the last term is of order ${\cal O}_r\left (\langle m\rangle
\left (b(m)\right )^4 \right )$. In what concerns 
$Q_0QE_1^2{\hat H}_0QQ_0$ one should use the perturbation theory up 
to $E_1$:
\begin{eqnarray}
Q_0QE_1^2{\hat H}_0QQ_0=Q_0E_1^2{\hat H}_0Q_0+
{\cal O}_r\left (\langle m\rangle
\left (b(m)\right )^3 \right )
\end{eqnarray}
so finally 
\begin{eqnarray*}
&&Q_0W^*{\hat H}_0WQ_0=Q_0{\hat H}_0Q_0+
Q_0({\hat H}_0E_1+E_1{\hat H}_0)Q_0\\ \nonumber
&&+\frac{1}{2}Q_0({\hat H}_0E_1^2+E_1^2{\hat H}_0)Q_0
+Q_0E_1{\hat H}_0E_1Q_0
+{\cal O}_r\left (\langle m\rangle \left (b(m)\right )^3\right ).
\end{eqnarray*}
 In the following we shall use the expansion $E_1={\hat E}_1+...+{\hat E}_N
+E_{N+1}$ for suitable $N$ and the estimate (\ref{QHE_j}), 
in such a way that the terms containing 
$E_{N+1}$ are small. It turns out that in order to estimate 
the third and the fourth terms it is sufficient to go up to 
$E_2$. When plugging this expansion in $Q_0{\hat H}_0E_1^2Q_0$
 and $Q_0E_1{\hat H}_0E_1Q_0$ several new terms will appear 
($Q_0{\hat H}_0{\hat E}_1E_2Q_0$, $Q_0{\hat E}_1{\hat H}_0E_2$ 
and $Q_0E_2{\hat H}_0E_2$ ). All of them are easily estimated in 
the same way as before and the result is:
\begin{eqnarray}\nonumber
Q_0{\hat H}_0E_1^2Q_0=Q_0{\hat H}_0{\hat E}_1^2Q_0
+{\cal O}_r\left (\langle m\rangle \left (b(m)\right )^3\right )\\
Q_0E_1{\hat H}_0E_1Q_0=Q_0{\hat E}_1{\hat H}_0{\hat E}_1
+{\cal O}_r\left (\langle m\rangle \left (b(m)\right )^3\right ).
\end{eqnarray}  
For the first term we have to go up to $E_3$ with the expansion. Using
(\ref{QHE_j}) and the identity (\ref{i1}) we arrive at 
\begin{eqnarray}
Q_0{\hat H}_0E_1Q_0=Q_0{\hat H}_0{\hat E}_2Q_0+
{\cal O}_r\left (\langle m\rangle \left (b(m)\right )^3\right ).
\end{eqnarray}
Moreover, using (\ref{i2}) we can replace $Q_0{\hat E}_2Q_0$ and write  
\begin{eqnarray}\nonumber
&&Q_0W^*H_0WQ_0=Q_0{\hat H}_0+Q_0(n_0-1)^2-
\frac{1}{2}Q_0({\hat H}_0{\hat E}_1^2+{\hat E}_1^2{\hat H}_0)Q_0
\\ \nonumber
&&+\frac{1}{2}(Q_0{\hat E}_1{\hat H}_0{\hat E}_1Q_0+
Q_0{\hat E}_1{\hat H}_0{\hat E}_1Q_0)+{\cal O}_r(\langle m\rangle \left (b(m)\right )^3)
\\ \nonumber
&&=Q_0H_0+\frac{1}{2}\left (Q_0[{\hat E}_1,{\hat H}_0]
{\hat E}_1Q_0+Q_0{\hat E}_1[{\hat H}_0,{\hat E}_1]\right )
+{\cal O}_r(\langle m\rangle \left (b(m)\right )^3)\\\nonumber
&&=Q_0H_0Q_0-\frac{1}{2}
\left\lbrace Q_0V{\hat E}_1Q_0+Q_0{\hat E}_1VQ_0 \right \rbrace\\
&&+{\cal O}_r\left (\langle m\rangle \left (b(m)\right )^3\right ).
\end{eqnarray}
In the last line we used (\ref{i3}). The lemma is now proved \qed   
\par

\vskip 0.2cm
\noindent
{\em Proof of Lemma 4.4:} 
Using (\ref{WQ_0}) and $Q=Q_0+E_1$ it results
\begin{eqnarray*}
Q_0W(t)^*VW(t)Q_0&=&Q_0VQ_0+Q_0(E_1V+VE_1)Q_0+Q_0E_1VE_1Q_0\\
&+&Q_0(Q_0+E_1)(VE_1^2F+FE_1^2V)(Q_0+E_1)Q_0\\
&+&Q_0(Q_0+E_1)FE_1(E_1VE_1)E_1F(Q_0+E_1)Q_0
\end{eqnarray*}
Let us first show that the last three terms are small. To see this
we need estimates on $E_1VE_1$ and $Q_0VE_1$. For $E_1VE_1$ one has
\begin{eqnarray}
E_1VE_1=-\frac{1}{2\pi ^2}\oint _{\gamma _{m}}
\oint _{\gamma _{m}}
dzdz'R_0(z,t)^{\frac{1}{2}}K(z,t)
\left (1+K(z,t)\right )^{-1}
\cdot \\ \nonumber
\cdot |R_0(z,t)|^{\frac{1}{2}}VR_0(z',t)^{\frac{1}{2}}K(z',t)
\left (1+K(z',t)\right )^{-1}
|R_0(z',t)|^{\frac{1}{2}}
\end{eqnarray}
Now  observe that
\begin{equation}
\sup _{t\in I_0}
\sup _{z\in \gamma _{m}}\sup _{z'\in \gamma _{m}}
\| |R_0(z,t)|^{\frac{1}{2}}VR_0(z',t)^{\frac{1}{2}} \|\leq
C_Vb(m).
\end{equation}
Then using the fact that $l(\gamma _{m})\sim \langle m\rangle $
\begin{eqnarray}\label{E_1VE_1}
\sup _{t\in I_0}\|E_1VE_1\|\leq  C_V\langle m\rangle
\left (b(m)\right )^3.
\end{eqnarray}
In what concerns $Q_0VE_1$ we have
\begin{eqnarray}\nonumber
\sup _{t\in I_0}\|Q_0VE_1\|&=&\frac{1}{2\pi}
\|\oint _{\gamma _{m}}\oint _{\gamma _{m}}dzdz'R_0(z,t)V
R_0(z',t)VR(z',t) \|\\\label{Q_0VE_1}
&\leq & C_V\langle m\rangle
\left (b(m)\right )^2
\end{eqnarray}
The estimates (\ref{E_1VE_1}) and (\ref{Q_0VE_1}) used in
$Q_0W(t)^*VW(t)Q_0$ suffices to prove that the last three terms are
of order ${\cal O}_r(\langle m\rangle
\left (b(m)\right )^3)$.
We arrived thus at
\begin{eqnarray*}
Q_0W^*VWQ_0&=&Q_0VQ_0+Q_0\left (VE_1+E_1V\right )Q_0\\
&+&{\cal O}_r(\langle m\rangle \left (b(m)\right )^3).
\end{eqnarray*}
As for the remaining term $Q_0E_1VQ_0$ we shall write it more
carefully by writing
\begin{equation}
Q_0E_1VQ_0=Q_0({\hat E}_1+E_2)VQ_0
\end{equation}
Finally, $Q_0E_2VQ_0$ is found to be of order
${\cal O}_r(\langle m\rangle \left (b(m)\right )^3) $ and the lemma
is finished.\qed
\noindent
\\
For ${\tilde H}_{{\rm eff},m}$ the computations and estimations 
are similar as above the difference appearing again due to the different integration contour. 
We can then conclude that 
\begin{eqnarray}\nonumber
{\tilde H}_{{\rm eff},m}(t)=
{\tilde Q}_{m,0}(H_0(t)+V){\tilde Q}_{m,0}+
\frac{1}{2}{\tilde Q}_{m,0}({\hat E}_{m,1}(t)V+V{\hat E}_{m,1}(t)){\tilde Q}_{m,0}+
\Delta {\tilde H}_{{\rm eff},m}(t),
\end{eqnarray} 
where ${\hat E}_{m,1}(t)$ is now the first order term from the pertubative expansion for 
${\tilde Q}_{m}(t)$ and the remainder $\Delta {\tilde H}_{{\rm eff},m}(t)$ is of the same 
order as in Theorem 4.2.   
\par

\setcounter{equation}{0} 
\section {The adiabatic evolution. The transition amplitudes  }
We shall use now the effective Hamiltonian obtained in the previous section to 
compute $U^A_{m}(t_1,t_0)$ and ${\tilde U}^A_{m}(t_2,t_1)$ 
when restricted to $Q_m(t_0)$ and ${\tilde Q}_m(t_1)$ respectively,
up to an error which remains small after taking the sum over $m$
from some sufficiently large $n_0$ to infinity. Remember that 
$t_0=-\frac{1}{4}$, $t_1=\frac{1}{4}$, $t_2=\frac{3}{4}$ and that 
$\{\varphi_{\alpha }(t_l)\}_{\alpha }$, $l=0,1,2$ are the canonical eigenbasis
for $H(t)$ (see (\ref{2.37})). The main result of this section is 
contained in: 
\vskip 0.2cm
\par
\noindent
{\sl{\bf Theorem 5.1   } Let $r>-\frac{1}{2}$, $m$ sufficiently large 
and define the transfer matrices
\begin{eqnarray}\label{calS1}
{\cal S}^m(t_1,t_0)&=&\left ( \begin{array}{cc}
{\cal S}^m_{2m-1,2m-2}(t_1,t_0) & {\cal S}^m_{2m-1,2m-1}(t_1,t_0) \\
{\cal S}^m_{2m-2,2m-2}(t_1,t_0) & {\cal S}^m_{2m-2,2m-1}(t_1,t_0)
\end{array}\right )\\\label{tildecalS1}
{\tilde{\cal S}}^m(t_2,t_1)&=&\left ( \begin{array}{cc}
{\tilde {\cal S}}^m_{2m,2m-1}(t_2,t_1) & {\tilde {\cal S}}^m_{2m,2m}(t_2,t_1) \\
{\tilde {\cal S}}^m_{2m-1,2m-1}(t_2,t_1) & {\tilde {\cal S}}^m_{2m-1,2m}(t_2,t_1)
\end{array}\right )
\end{eqnarray}

with
\begin{eqnarray}\label{calS}
{\cal S}_{i,j}^m(t_1,t_0)&:=&\langle\varphi_i(t_1),
U^A_{m}(t_1,t_0)
\varphi_j(t_0)\rangle ,\qquad i,j =2m-2,2m-1\\\label{caltildeS}
{\tilde {\cal S}}_{i,j}^m(t_2,t_1)&:=&\langle\varphi_i(t_2),
{\tilde U}^A_{m}(t_2,t_1)
\varphi_j(t_1)\rangle,\qquad i,j =2m-1,2m.
\end{eqnarray}
Then 
\begin{eqnarray}
{\cal S}^m(t_1,t_0)&=& e^{-i\theta_{m}(t_1,t_0)}\left ( \begin{array}{cc} 
\alpha _{2m-2}(t_1,t_0) & -\overline  {\beta}_{2m-2}(t_1,t_0)\\
\beta_{2m-2}(t_1,t_0) &\alpha '_{2m-2}(t_1,t_0)
\end{array}\right )+{\cal R}_{{\rm eff},m},\\
{\tilde {\cal S}}^m(t_2,t_1)&=&e^{-i{\tilde\theta}_{m}(t_2,t_1)}\left ( \begin{array}{cc}
{\tilde\alpha } _{2m-1}(t_2,t_1) & -\overline  {\tilde{\beta}}_{2m-1}(t_2,t_1)\\
{\tilde\beta }_{2m-1}(t_2,t_1) &{\tilde\alpha} ' _{2m-1}(t_2,t_1)
\end{array}\right )+{\cal R}_{{\rm eff},m},
\end{eqnarray}
with:
\begin{equation}\label{erorad}
{\cal R}_{{\rm eff},m}=
{\cal O}_r\left (
\frac{ \log ^8\langle m  \rangle}{\langle m  \rangle^{\min  \{2+8r,\frac{3}{2}+5r,\frac{3}{2}+r,2 \}} }
\right )\qquad {\rm as}\,\, m\to\infty 
\end{equation} 
and
the following notations (we list below only the quantities related to 
${\cal S}^m$; the ones corresponding to ${\tilde {\cal S}}^m$ are discussed at the end of 
this section)
\begin{eqnarray}\label{a}
\alpha _{2m-2}(t_1,t_0)&:=&1-\frac{|{\hat V}(2(m-1))|^2}{m-1}c_{2,m}(t_1,t_0)
-i\omega_{1,m}(t_1,t_0),\\
\alpha '_{2m-2}(t_1,t_0)&:=&1-\frac{|{\hat V}(2(m-1))|^2}{m-1}c_{2,m}(t_1,t_0)
-i\omega_{2,m}(t_1,t_0),
\\\label{b}
\beta_{2m-2}(t_1,t_0)&:=&\frac{ \overline {{\hat V}(2(m-1))}}{\sqrt{m-1}}c_{1,m}(t_1,t_0)
-i\lambda _{m}(t_1,t_0)-\frac{ 2\overline{{\hat V}(2(m-1))}  }{m-1},\\
c_{1,m}(t_1,t_0)&:=&-{(i+1)\frac{\sqrt{\pi}}{2}}\,e^{i\frac{m-1}{8}}
\,\erf\left (\frac{i+1}{4}\sqrt{m-1}\right ),\\
c_{2,m}(t_1,t_0)&:=&\frac{\pi}{4}\,\left |\erf\left (\frac{i-1}{4}
\sqrt{m-1}\right )\right |^2,
\end{eqnarray}
\begin{eqnarray}\nonumber
\theta _{m}(t_1,t_0)&:=&\int_{t_0}^{t_1}ds\left ( (m-1)^2+s^2\right ),\\    
\lambda _{m}(t_1,t_0)&:=&\int_{t_0}^{t_1}ds{\overline \Delta }_{m} (s)
e^{-2i(m-1)(s^2-t_0^2)},  
\end{eqnarray}
\begin{eqnarray}\nonumber
\omega_{1,m}(t_1,t_0)&:=&
\int_{t_0}^{t_1}ds\frac{1}{2}\langle\varphi_{m-1,0},
(V{\hat E}_1(t)+{\hat E}_1(t)V)\varphi_{m-1,0}  \rangle\\\label{o1}
&=:&\int_{t_0}^{t_1}ds\gamma _{1,m}(s)\\\nonumber
\omega_{2,m}(t_1,t_0)&:=&
\int_{t_0}^{t_1}ds\frac{1}{2}\langle\varphi_{-(m-1),0},
(V{\hat E}_1(t)+{\hat E}_1(t)V)\varphi_{-(m-1),0}  \rangle\\\label{o2}
&=:&\int_{t_0}^{t_1}ds\gamma _{2,m}(s),\\\label{deltam}
\Delta_{m}(t)&:=&\frac{1}{2}\langle\varphi_{m-1,0},
(V{\hat E}_1(t)+{\hat E}_1(t)V)\varphi_{-(m-1),0}  \rangle.
\end{eqnarray}
Here $\erf(z)$ is the error function (see \cite{AS}):
\begin{equation}
\erf(z):={\frac{2}{\sqrt{\pi}}}\int_0^z e^{-t^2}dt.
\end{equation}
}
\vskip 0.2 cm
\noindent
{\bf Remark:} Although the above estimates are valid for $r>-\frac{1}{2}$ the 
bound on the error terms gets small as $m$ tends to infinity only for $r>-\frac{1}{4}$
and is summable with respect to $m$ only for $r>-\frac{1}{10}$ (these facts can
be noticed from the behaviour of the error term ${\cal R}_{{\rm eff},m}$).

\vskip 0.2 cm
\noindent
{\em Proof:}
We shall consider only $S^m_{i,j}$; the proof for ${\tilde S }^m_{i,j}$ is similar.
From (\ref{2.37}) and (\ref{4.12}):  
\begin{equation}\label{5.14}
S^m_{i,j}(t_1,t_0)=
\sum_{k,l=2m-2,2m-1}\overline {c}_{k,i}^m(t_1)
{\cal U}_{{\rm eff},m;k,l} (t_1,t_0)c_{l,j}^m(t_0),
\end{equation}
where  
$\varphi_{\alpha}^0(t_{\beta })=\varphi_{n_{\alpha},0}(t_{\beta })$ (see (\ref{labelling1}))
 and we introduced
the coefficients 
\begin{equation}
c^m_{i,k}(t_p):=\langle\varphi_i^0(t_p),W_{m}(t_p)^*N_k(t_p)\varphi_k^0(t_p)
\rangle ,\qquad p=0,1 
\end{equation}
and 
\begin{equation}
{\cal U}_{{\rm eff},m;k,l}(t_1,t_0):=\langle\varphi_k^0(t_1),{\cal U}_{{\rm eff},m}(t_1,t_0)
\varphi_l^0 (t_0)\rangle .
\end{equation}
The estimation of the coefficients $c^m_{i,k}(t_p)$ is easy and reads:
\vskip 0.2cm
\par
\noindent
{\sl{\bf Lemma 5.2: } For $m$ large enough and $r>-\frac{1}{2}$ one has:
\begin{equation}\label{5.17}
c_{\alpha ,\beta }^m(t_p)=\delta _{\alpha ,\beta } +
\frac{\langle\varphi_{\alpha}^0(t_p),V\varphi_{\beta}^0(t_p)\rangle}  
{E_{\beta}^0(t_p)-E_{\alpha}^0(t_p)}(1-\delta_{\alpha ,\beta}) +{\cal O}_r(b^2(m)),
\qquad {\rm as}\,\, m\to\infty.
\end{equation}
}
\vskip 0.2cm
The estimation of ${\cal U}_{{\rm eff},m}$ is more involved and the result
is summarised in:
\vskip 0.2cm
\par
\noindent
{\sl{\bf Lemma 5.3:}  For $m$ large enough and $r>-\frac{1}{2}$ the effective evolution 
is given by:
\begin{equation}\label{5.18}
{\cal U}_{{\rm eff},m;i,j}(t_1,t_0)={\cal S}^m_{ij}(t_1,t_0) +e^{-i\theta_m(t_1,t_0)}v_{m}
+{\cal R}_{{\rm eff},m}, \qquad {\rm as}\,\, m\to\infty.
\end{equation}
with 
\begin{eqnarray}
v_m:=\left ( \begin{array}{cc}
0 &  -\frac { 2{\hat V}(2(m-1))}{m-1} \\
 \frac {2\overline {{\hat V}(2(m-1))}}{m-1}      &0
\end{array}\right ).
\end{eqnarray}

}
\noindent
The proof of Thm. 5.1 requires only straightforward calculations 
using (\ref{5.14}), (\ref{5.17}) and (\ref{5.18}). 

\vskip 0.2 cm
\noindent
{\em Proof of Lemma 5.2:} Write (see (\ref{4.21}) and (\ref{4.22})) 
\begin{equation}
Q_m(t_p)-Q_{m}^0(t_p)={\hat E}_{m,1}(t_p)+E_{m,2}(t_p),
\end{equation}
and in the same way 
\begin{eqnarray}\nonumber
P_{\alpha}(t_p)-P_{\alpha}^0(t_p)&=&-\frac{i}{2\pi}\int_{C_{\alpha }}
dz R_0(z,t_p)VR_0(z,t_p)\\\nonumber
&+&\frac{i}{2\pi}\int_{C_{\alpha }} dzR_0(z,t_p)
VR(z,t_p)VR_0(z,t_p)\\
&=:&{\hat e}_{\alpha ,1}(t_p)+e_{\alpha ,2}(t_p),\qquad p=0,1
\end{eqnarray}
with $C_{2m-2}=\Gamma_{m-1}\cup {\tilde\Gamma}_{m-1} $ and
$C_{2m-1}={\tilde\Gamma}_{m-1}\cup\Gamma_{m}$, i.e $C_{\alpha}$ is a countour 
which contains only one eigenvalue of $H_0(t_p)$ namely $E_{\alpha}^0(t_p)$. 
By Lemma 2.3 (see also Section 4)one has 
\begin{eqnarray}\label{*}
\|{\hat E}_{m,1}(t_p)\|+\|{\hat e}_{\alpha ,1}(t_p)\|\leq C_Vb(m),\\\label{**}
\|E_{m,2}(t_p)\|+\|e_{\alpha ,2}(t_p)\|\leq C_Vb(m)^2.
\end{eqnarray}
Then using the expansion in the Sz-Nag\"y formula, (\ref{*}), (\ref{**})
and the fact that for $\alpha =2m-2,2m-1$ we have 
$Q_{m,0}P_{\alpha}^0(t_p)=P_{\alpha}^0(t_p)Q_{m,0}=P_{\alpha}^0(t_p) $ and 
$P_{\alpha}^0(t_p)\varphi_{\alpha}^0(t_p)=\varphi_{\alpha}^0(t_p)$ 
it follows that up to errors of order
${\cal O}_rb(m)^2$ 
\begin{equation}
c_{\alpha ,\beta }^m(t_p)=\langle \varphi_{\alpha}^0(t_p),
(1+{\hat E}_{m,1}(t_p)+{\hat e}_{\beta ,1}(t_p))\varphi_{\beta}^0(t_p)\rangle.
\end{equation}
Since $Q_{m,0}{\hat E}_{m,1}(t_p)Q_{m,0}=P_{\beta}^0(t_p){\hat e}_{\beta ,1}(t_p)
P_{\beta}^0(t_p)=0$ one gets finally
\begin{equation}\label{5.25}
|c_{\alpha ,\beta }^m(t_p)-\delta_{\alpha,\beta}-
\langle \varphi_{\alpha}^0(t_p),{\hat e}_{\beta ,1}(t_p)\varphi_{\beta}^0(t_p)
(1-\delta_{\alpha,\beta})\rangle|
\leq  C_Vb(m)^2
\end{equation}
and computing the third term in the LHS of (\ref{5.25}) 
by residues theorem we obtain the needed 
result.\qed 

\par
\vskip 0.2 cm
\noindent
{\em Proof of Lemma 5.3:}
 First remark that due 
to Theorem 4.1 one can write 
\begin{equation}
H_{{\rm eff},m}(t)=h_{{\rm eff},m}(t)+\Delta H_{{\rm eff},m}(t),
\end{equation}
where 
\begin{equation}
h_{{\rm eff},m}(t)=Q_{m,0}\left (H_0(t)+V\right )Q_{m,0}+
\frac{1}{2}Q_{m,0}({\hat E}_{m,1}(t)V+V{\hat E}_{m,1}(t))Q_{m,0}.
\end{equation}
Denote then by ${\cal U}_{{\rm eff },m}^0(t,t_0)$ the evolution associated to 
$h_{{\rm eff},m}(t)$ having the equation of motion 
\begin{equation}
i\frac{d}{dt}{\cal U}_{{\rm eff },m}^0(t,t_0)=h_{{\rm eff},m}(t)
{\cal U}_{{\rm eff },m}^0(t,t_0),
\quad {\cal U}_{{\rm eff },m}^0(t_0,t_0)=1.
\end{equation} 
The usual estimation for 
$\Omega _{{\rm eff },m}(t,t_0):={\cal U}_{{\rm eff },m}^0(t,t_0)^*
{\cal U}_{{\rm eff },m}(t,t_0)$ gives 
\begin{eqnarray*}
\sup_{t\in I_0}\|\Omega_{{\rm eff },m} (t,t_0)-1\|&\leq& 
\sup_{t\in I_0}\|i\int_{t_0}^tds{\cal U}_{{\rm eff },m}^0(s,t_0)^*
\Delta H_{{\rm eff},m}(s){\cal U}_{{\rm eff },m}^0(s,t_0)\Omega _{{\rm eff },m}
(s,t_0)\| \nonumber \\
&\leq&\frac{1}{2}\sup_{t \in I_0}\|\Delta H_{{\rm eff},m}(s)
\|\leq C_V\langle m  \rangle b(m)^3. 
\end{eqnarray*}
Hence
\begin{equation}\label{estim1}
{\cal U}_{{\rm eff },m}(t,t_0)={\cal U}_{{\rm eff },m}^0(t,t_0)+
{\cal O}_r(\langle m  \rangle b(m)^3),\qquad {\rm as}\,\, m\to\infty .
\end{equation}
We continue the computation of ${\cal U}_{{\rm eff },m}^0(t,t_0)$ by 
further decomposing $h_{{\rm eff},m}(t)$ 
into a "free" part 
\begin{equation}
H_{P,m}(t):=Q_{m,0}\left (H_0(t)+V\right )Q_{m,0}
\end{equation}
and a perturbation
\begin{equation}\label{defBm}
B_m(t):=\frac{1}{2}Q_{m,0}\left ({\hat E}_{m,1}(t)V+V{\hat E}_{m,1}(t)\right )Q_{m,0}.
\end{equation}
Remark first that $H_{P,m}(t)$ has a simple matrix form in the eigenbasis of $H_0(t)$ 
 that can be explicitely
written taking into account that $Q_{m,0}=P_{m-1,0}+P_{-(m-1),0}$.
Secondly one notes that it is possible to eliminate a diagonal term from 
$H_{P,m}(t)$ by writing 
\begin{equation}\label{wh}
{\cal U}_{{\rm eff },m}^0(t_1,t_0):=\exp\left (-i\int_{t_0}^{t_1}\left ((m-1)^2+s^2\right )ds\right )
\widehat {\cal U}_{{\rm eff },m}(t_1,t_0),
\end{equation}
where  $\widehat {\cal U}_{{\rm eff },m}(t,t_0)$ satisfies the equation
\begin{equation}
i\frac{d}{dt}\widehat {\cal U}_{{\rm eff },m}(t,t_0)=
\left (\widehat H_{P,m}(t)+B_{m}(t)\right )\widehat {\cal U}_{{\rm eff },m}(t,t_0),
\quad \widehat {\cal U}_{{\rm eff },m}(t_0,t_0)=1,
\end{equation}
and
\begin{equation}
\widehat H_{P,m}(t)=\left (\begin{array}{cc} 2 (m-1)t& {\hat V}(2(m-1))\\
                       \overline {{\hat V}(2(m-1))} & -2(m-1)t\end{array}\right ).
\end{equation}
Now let $\widehat {\cal U}_{P,m}(t,t_0)$ be the evolution generated by 
$\widehat H_{P,m}(t)$ and
\begin{equation}\label{omegaPm}
\widehat\Omega _{P,m}(t_1,t_0)=\widehat {\cal U}_{P,m}(t_1,t_0)^*
\widehat {\cal U}_{m,{\rm eff} }(t_1,t_0).
\end{equation}
We write the Dyson expansion for $\widehat\Omega _{P,m}$
with the remainder of order 2:
\begin{eqnarray}\nonumber
\widehat\Omega _{P,m}(t_1,t_0)&=&1+(-i)\int_{t_0}^{t_1}ds_1 B_{m}^{{\rm int}}(s_1)
\\ 
&+&(-i)^2\int_{t_0}^{t_1}ds_1B_{m}^{{\rm int}}(s_1)
\int_{t_0}^{s_1}ds_2B_{m}^{{\rm int}}(s_2)\widehat\Omega _{P,m}(s_2,t_0), 
\end{eqnarray}
where 
$B_{m}^{{\rm int}}(t)=\widehat {\cal U}_{P,m}(t,t_0)^*
B_{m}(t)\widehat {\cal U}_{P,m}(t,t_0)$.
Observe first that $\sup_{t\in I_0}B_m(t)={\cal O}_r(\langle m\rangle b(m)^2)$
as $m\to\infty$
(this follows from the estimate on $Q_{m,0}V{\hat E}_1(t)$ which is the same 
as the one from Eq.(\ref{Q_0VE_1})). Then
\begin{eqnarray}\nonumber
&&\sup_{t\in I_0}\|
\int_{t_0}^{t_1}ds_1B_{m}^{{\rm int}}(s_1) 
\int_{t_0}^{s_1}ds_2B_{m}^{{\rm int}}(s_2)\widehat\Omega _{P,m}(s_2,t_0) \|
\leq \frac{1}{4}\left (\sup_{t\in I_0}\|B_{m}(t)\|\right )^2\\\label{5.36} 
&&= {\cal O}_r\left ( \left (\langle m\rangle b(m)^2\right )^2\right ),\qquad {\rm as}\,\,
 m\to\infty .
\end{eqnarray}
In conclusion, at this level of approximation from (\ref{omegaPm}) we have 
\begin{equation}\label{Bint}
\widehat {\cal U}_{{\rm eff },m}(t_1,t_0)=\widehat {\cal U}_{P,m}(t_1,t_0)
\left [1-i\int_{t_0}^{t_1}dsB_{m}^{{\rm int}}(s) \right ]+
{\cal O}_r\left (\left (\langle m\rangle b(m)^2\right )^2\right ), \qquad {\rm as}\,\, 
m\to\infty .
\end{equation}
Thus we have reduced the problem at hand to the computation of
$\widehat {\cal U}_{P,m}(t_1,t_0) $ and of the contribution of $B_{m}^{{\rm int}}$.
Now we write $\widehat H_{P,m}(t)$ in the form             
$\widehat H_{P,m}(t)=\widehat H_{P,m}^0(t)+\widehat V_P$,
where 
\begin{eqnarray*}
\widehat H_{P,m}^0(t)=\left (\begin{array}{cc} 2(m-1)t& 0\\
                       0 & -2(m-1)t\end{array}\right ), \quad
\widehat V_P=\left (\begin{array}{cc}0&  {\hat V}(2(m-1))\\
             \overline {{\hat V}(2(m-1))} & 0\end{array}\right ). 
\end{eqnarray*}
Let $\widehat {\cal U}_{P,m}^0$ be the evolution
 generated by $\widehat H_{P,m}^0 $:
\begin{eqnarray}\label{uPmzero}
\widehat {\cal U}_{P,m}^0(t,t_0)
&=&
\left (\begin{array}{cc} e^{-i(m-1)(t^2-t_0^2)}&0\\
             0& e^{i(m-1)(t^2-t_0^2)}\end{array}\right ).
\end{eqnarray}                        
By defining all the "effective"
Hamiltonians above we ended with a time-independent perturbation
$\widehat V_P$ (which is precisely the one considered in the physical literature 
\cite{Ao,Br}). Writing the Dyson expansion with remainder of order 3 
for 
$\widehat\Omega _{P,m}^0(t_1,t_0)=\widehat {\cal U}_{P,m}^0(t_1,t_0)^*
\widehat {\cal U}_{P,m}(t_1,t_0)$:
\begin{eqnarray}\nonumber   
\widehat\Omega _{P,m}^0(t_1,t_0)
\nonumber
&=&1+(-i)\int_{t_0}^{t_1}ds_1\widehat V_P(s_1)
+(-i)^2\int_{t_0}^{t_1}ds_1\widehat V_P(s_1)
\int_{t_0}^{s_1}ds_2\widehat V_P(s_2) \\\nonumber 
&+&(-i)^3\int_{t_0}^{t_1}ds_1\widehat V_P(s_1)
\int_{t_0}^{s_1}ds_2\widehat V_P(s_2)
\int_{t_0}^{s_2}ds_3\widehat V_P(s_3)\widehat\Omega _{P,m}^0(s_3,t_0)\\\label{R3}
&:=&1+\widehat\Omega_{P,m}^{0,(1)}(t_1,t_0)+\widehat\Omega_{P,m}^{0,(2)}(t_1,t_0)
+{\cal R}_3,
\end{eqnarray}
where 
$\widehat V_P(t)=\widehat {\cal U}_{P,m}^0(t,t_0)^*\widehat V_P
\widehat {\cal U}_{P,m}^0(t,t_0)$.
We compute explicitely the first and second order  
terms while the 
remainder ${\cal R}_3$ is estimated using Theorem 7.3.2 from Appendix 7.3. 
That is, we use (\ref{reste})
for $p=3$ which gives, with $m$ large enough 
\begin{equation}\label{resteR3}
\|{\cal R}_3\|\leq C_{m_0}\|V\|_r^3\frac {\log ^2 \langle m \rangle }
{\langle m  \rangle^{\frac{3}{2}+3r}}. 
\end{equation}
As for the first two terms in the Dyson expansion by direct computation we have:
\begin{equation}\label{omega}
\sum_{k=0}^2\widehat\Omega_{P,m}^{0,(k)}(t_1,t_0)=
\left (\begin{array}{cc} 1-\frac{|{\hat V}(2(m-1))|^2}{m-1}c_{2,m}(t_1,t_0)&
{-\frac{ {\hat V}(2(m-1))}{ \sqrt{m-1}}}\,\overline{c_{1,m}(t_1,t_0)}\\
{\frac{ \overline {{\hat V}(2(m-1))}}{\sqrt{m-1}}}\,c_{1,m}(t_1,t_0)
                & 1-\frac{|{\hat V}(2(m-1))|^2}{m-1}c_{2,m}(t_1,t_0)
\end{array}\right ),
\end{equation}
with $c_{1,m}(t_1,t_0)$ and $c_{2,m}(t_1,t_0)$ as introduced in Thm. 5.1.  
\par

The expression (\ref{omega}) together with the estimate (\ref{resteR3}) allow to compute 
approximately 
$\widehat {\cal U}_{P,m}$ 
with the help of the following formula for  
$B_{m}(t)$ (see (\ref{defBm}) and (\ref{o1})-(\ref{deltam})): 
\begin{equation}
B_{m}(t)=\left ( \begin{array}{cc} \gamma _{1,m}(t) &\Delta_{m} (t)\\
                   {\overline \Delta }_{m} (t)  &\gamma _{2,m}(t)\end{array}\right ).
\end{equation}
Later on we shall need the the following expansion of $B_m^{{\rm int}}$:
\begin{eqnarray*}
B_m^{{\rm int }}(t)&=&\widehat {\cal U}_{P,m}(t,t_0)^*
B_{m}(t)\widehat {\cal U}_{P,m}(t,t_0)\\
&=&\left (\widehat {\cal U}^0_{P,m}(t,t_0)
\left (1+(-i)\int_{t_0}^{t}ds{\hat V}_P(s){\widehat\Omega}^0_{P,m}(s,t_0)  \right )  \right )^*
B_m(s)\widehat {\cal U}^0_{P,m}(t,t_0)\\
&\times &\left (1+(-i)\int_{t_0}^{t}ds{\hat V}_P(s){\widehat\Omega}^0_{P,m}(s,t_0)  \right ).
\end{eqnarray*}
The term $(-i)\int_{t_0}^{t}ds{\hat V}_P(s){\widehat\Omega}^0_{P,m}(s,t_0)$ is actually 
the remainder of order 1 from the Dyson expansion of ${\widehat\Omega}^0_{P,m}(s,t_0)$.
Consequently we denote this quantity ${\cal R}_1$  
and estimate it by (\ref{reste})(for $m$ large enough) as
\begin{equation}\label{R1}
\|{\cal R}_1\|\leq C_{m_0}\|V\|_r\frac {\log \langle m \rangle }
{ \langle m  \rangle^{\frac{1}{2}+r}}.
\end{equation}
Using that $B_m(t)={\cal O}_r((b(m))^2\langle m  \rangle )$
and (\ref{R1}) we have  
\begin{equation}\label{estim2}
B_{m}^{{\rm int}}(t)=\widehat {\cal U}_{P,m}^0(t,t_0)^*B_{m}(t)\widehat {\cal U}_{P,m}^0(t,t_0)
+{\cal O}_r\left (\frac{\log^5 \langle m \rangle }
{\langle m\rangle
^{ \min \{\frac{3}{2}+r,\frac{3}{2}+5r \} } }             \right ), {\rm as }\,\, m\to\infty .
\end{equation}
To write down ${\cal U}_{{\rm eff },m}(t_1,t_0)$ we use (\ref{estim1}), (\ref{wh}),
 (\ref{omegaPm}), (\ref{5.36}),(\ref{uPmzero}), (\ref{R3}), (\ref{omega}) and 
(\ref{estim2}): 
\begin{eqnarray}\nonumber
{\cal U}_{{\rm eff },m}(t_1,t_0)&=&
e^{-i\theta_m(t_1,t_0)}
\left (\begin{array}{cc} 1-\frac {|{\hat V}(2(m-1))|^2}{m-1}c_{2,m}(t_1,t_0)
& -\frac{  {\hat V}(2(m-1))} {    \sqrt {m-1}}
 \overline {c_{1,m}(t_1,t_0)}  \\
\frac{ \overline {{\hat V}(2(m-1))}}{\sqrt {m-1}}
\overline c_{1,m}(t_1,t_0)
&1-\frac {|{\hat V}(2(m-1))|^2}{m-1}c_{2,m}(t_1,t_0)\end{array}\right )\\\label{ueff}
&-&i\left( \begin{array}{cc} \omega _{1,m} &\overline{\lambda _{m}} \\
\lambda _{m}&\omega _{2,m}\end{array}\right )+{\cal R}_{{\rm eff },m}. 
\end{eqnarray}
The remainder ${\cal R}_{{\rm eff },m}$ takes into account all the errors
involved during the computation of the effective evolution in the subspace $Q_m{\cal H}$:   
\begin{eqnarray}\nonumber
\|{\cal R}_{{\rm eff },m}\|&\leq &
2\|{\cal R}_1\|\cdot \|B_m\|+\|{\cal R}_3\|+
\|B_m)\|^2+\|\Delta H_{{\rm eff},m}\|\\
&=& {\cal O}_r\left (
\frac{ \log ^8\langle m  \rangle}{\langle m  \rangle^{\min  \{2+8r,\frac{3}{2}+5r,\frac{3}{2}+r,2 \}} }
\right ). 
\end{eqnarray}
Making the necessary identifications in (\ref{ueff}) one finds that this expression coincides with the one
 given in Lemma 5.3 
which is now proven. \qed
\vsth
\nid
{\bf Remark 5.4:} Let us justify why we kept explicitely the matrix elements of 
$B_m(t)$ in the 
effective evolution.  
 First we point out, without givind details, that one can obtain sharp estimates 
on $\Delta_{m}$ and $\gamma $'s as in Eqs.(10) and (11)
from \cite{ABDN}, namely ($j=1,2$)
\begin{eqnarray}\label{gamma}
\sup_{t\in I_0}|\Delta_{m}(t)|\leq \frac{C\|V\|^2_r}{\langle m \rangle^{1+r} },\qquad
\sup_{t\in I_0}|\gamma _{j,m}(t)|\leq \frac{C\|V\|^2_r}{\langle m \rangle^{\min \{2,1+2r\}} }.
\end{eqnarray}
Secondly, observe that from these estimates one cannot conclude that the diagonal contribution 
of $B_m(t)$ is smaller that the second terms from the Landau-Zener 
coefficients $\alpha_{2m-2}, \alpha'_{2m-2}$.        

\noindent
\\
To obtain ${\tilde {\cal U}} _{{\rm eff },m}(t_2,t_1)$ we have to follow the same steps 
as above, using the effective Hamiltonian ${\tilde H}_{{\rm eff},m}$.
 Without giving explicit calculations we summarise the results and 
the notations. Up to errors of order 
${\cal O}_r(\langle m \rangle ^{\min \{2,2+8r\} })$ as $m\to\infty $: 
\begin{eqnarray}\nonumber
{\tilde {\cal U}} _{{\rm eff },m}(t_2,t_1)&=&e^{-i{\tilde\theta}_m(t_2,t_1)}
\left ( \begin{array}{cc} 1-\frac {|{\hat V}(2m-1)|^2}{m-\frac{1}{2}}
{\tilde c}_{2,m}(t_2,t_1)
& -\frac{ {\hat V}(2m-1)}{\sqrt {m-\frac{1}{2}}}
\overline{{\tilde c}}_{1,m}(t_2,t_1)
)  \\
\frac{ \overline {{\hat V}(2m-1)}}{\sqrt {m-\frac{1}{2}}}
{\tilde c}_{1,m}(t_2,t_1)
&1-\frac {|{\hat V}(2m-1)|^2}{m+\frac{1}{2}}{\tilde c}_{2,m}
(t_2,t_1)\end{array}\right )\\
 &-&ie^{-i{\tilde\theta}_m(t_2,t_1)} 
 \left (\begin{array}{cc}{\tilde\omega} _{1,m}(t_2,t_1)
  &\overline{{\tilde\lambda} _{m}(t_2,t_1)} \\
  {\tilde\lambda}_{m}(t_2,t_1)&
  {\tilde\omega} _{2,m}(t_2,t_1) \end{array}\right )  
  \end{eqnarray}
the phase ${\tilde\theta}_m$ coeficients ${\tilde c}_i(m)$ being related to the ones of 
$ {\cal U}_{{\rm eff },m}(t_1,t_0)$ by the relations
\begin{eqnarray}
{\tilde\theta}_m(t_2,t_1)&=&\theta_{m+\frac{1}{2}}(t_1,t_0)\\
{\tilde c}_{i,m}(t_2,t_1)&=&c_{i,m+\frac{1}{2}}(t_1,t_0)
,\qquad i=1,2.
\end{eqnarray}
${\tilde\lambda} _{m}(t_2,t_1)$ and ${\tilde\gamma} _{1,m}(t)$
are to be computed in the same way as $\lambda_{m}(t_1,t_0)$ 
and $\gamma_{m}(t_1,t_0)$
(replace $\varphi_{-(m-1),0}$ with $\varphi_{-m,0}$ and ${\hat E}_{m,1}(t)$ with 
${\tilde {\hat E}}_{m,1}(t)$ associated to ${\tilde Q}_m(t)$ ). Finally:  
\begin{eqnarray}\label{atilde}
{\tilde\alpha}_{2m-1}(t_2,t_1)&:=&1-\frac {|{\hat V}(2m-1)|^2}{m-\frac{1}{2}}
{\tilde c}_{2,m}(t_2,t_1)
-i{\tilde\omega}_{1,m}(t_2,t_1)
 \\\label{btilde}
 {\tilde\beta}_{2m-1}&:=&\frac{ \overline {{\hat V}(2m-1) }}{\sqrt {m-\frac{1}{2}}}
{\tilde c}_{1,m}(t_2,t_1)-i{\tilde\lambda}_{m}(t_2,t_1).
\end{eqnarray}
The proof of Theorem 5.1 is finished. \qed
\par

\setcounter{equation}{0}

\section {The long time behaviour. Spectral properties}
\par
The last part of this work is concerned with the long time behaviour of the 
Bloch electron. Suppose that at the time $t_0=-\frac{1}{4}$ the electron is in a 
given band (say $E_{2n_0-2}(t_0)$ for $n_0$ fixed) and  
its wavefunction $\psi(t_0)=\varphi_{2n_0-2}(t_0)$ (let us recall that 
$\varphi_i(t)$ is the eigenfunction of $H(t)$ which corresponds to the 
eigenvalue $E_i(t)$). 
 We are interested in obtaining some quantitative information    
 about $\psi(t_N)=U(t_N,t_0)\psi(t_0)$ 
where $t_N=t_0+\frac{N}{2}$, particular attention being payed to the case
when $N$ 
goes to infinity. 

To answer these questions we proceed as follows: the 
complete evolution $U(t_N,t_0)$ is written as a product of one-period evolutions
$U(t_{l}+1,t_l)$ each of them being then reduced to the first period 
$I_0\cup I_1$ by  using (\ref{2.22}).  
 Then we use Theorem 3.1 to approximate the one-period evolution by 
the adiabatic evolutions  
which were written explicitely (in suitable subspaces) in Theorem 5.1. 
As pointed out by Ao it is of particular interest to establish 
how far the electron goes in the energy space, otherwise stated, to 
say up to what band it is accelerated by the electric field. Because  
in the neighbour band approximation the electron can jump during one
half period only up to the next band it is clear that after 
$N$ half periods the uppermost reachable band is 
$E_{2n_0-2+N}(t_N)$. Its eigenfunction $ \varphi_{2n_0-2+N}(t_N)$ is related
to $\varphi_{2n_0-2+N}(t_0)$ by the shift operator $T$ (see (\ref{2.38}) ).  
The long time behaviour of the Bloch electron is described in the following 
theorem:
\par
\vskip 0.2cm
\noindent {\sl {\bf Theorem 6.1:} Let $r>0$ and $n_0$ sufficiently
large. Define the so-called 
propagating front \cite{Ao}:    
\begin{equation}\label{6.1}
{\cal P}(N):=|\langle \varphi_{2n_0-2+N}(t_N),U(t_N,t_0)
\varphi_{2n_0-2}(t_0)\rangle |^2.
\end{equation} 
Then 
\begin{equation}\label{propagf}
\liminf_{N\to\infty} {\cal P}(N)
=\exp \left\lbrace -2\sum_{l=0}^{\infty}
\frac {|{\hat V}(2n_0+l-2)|^2}{n_0+\frac{l}{2}-1}c_{2,n_0+\frac{l}{2}}(t_1,t_0)
 \right \rbrace\times
 \left (1+{\cal R}\right ),
\end{equation}
with 
\begin{equation}
|{\cal R}|\leq 
 \sum_{l=0}^{\infty}
\frac {C_V}{\langle n_0+l-\frac{1}{2}\rangle ^{1+2r}}+
\frac{ C_V\log ^8\langle n_0 \rangle }{\langle n_0
\rangle^{\min  \{1+8r,\frac{1}{2}+5r,\frac{1}{2}+r,1 \}} }\exp \left (\sum_{l=0}^{\infty}
\frac {C_V}{\langle n_0+\frac{l}{2}-1\rangle ^{1+2r} } \right ).
\end{equation}}
\vskip 0.2cm
\noindent
\nid
As one may guess we need first a more explicit form for the scalar
product in Eq.(\ref{6.1}). Consequently a preliminar result is 
\vskip 0.2cm
\noindent {\sl {\bf Lemma 6.2:} 
 For $r>-\frac{1}{10}$, $n_0$ sufficiently large
 and $N$ even one gets:
\begin{eqnarray}\nonumber 
&&\langle \varphi_{2n_0+N-2}(t_N),U(t_N,t_0)
\varphi_{2n_0-2}(t_0)\rangle
\\\label{prod}
&=&\prod_{l=0}^{\frac{N}{2}-1}e^{-i({\tilde\theta}_{n_0+l}(t_2,t_1)+\theta_{n_0+l}(t_1,t_0))}
{\tilde\alpha}_{2(n_0+l)-1}(t_2,t_1)
\alpha_{2(n_0+l)-2}(t_1,t_0)+{\cal R}(N),
\end{eqnarray}
where the remainder ${\cal R}(N)$ satisfy the estimate 
\begin{eqnarray}
|{\cal R}(N)|\leq 
\frac{ C_V\log ^8\langle n_0 \rangle }{\langle n_0  \rangle^{\min  \{1+8r,\frac{1}{2}+5r,\frac{1}{2}+r,1 \}} }
\end{eqnarray}
A similar expression holds for $N$ odd.}
\skip 0.2cm
\\
\vsth
\noindent
{\em Proof:}
The idea is to factorise $U(t_N,t_0)$ into one-period evolutions and to use the adiabatic
theorem for a suitable index $m$ of the two-dimensional projector $Q_m(t)$ or
${\tilde Q}_m(t)$. By the telescoping sum rule:
$$
\prod_{l=0}^na_l-\prod_{l=0}^nb_l=\sum_{l'=0}^n\prod_{l=l'+1}^na_l(a_{l'}-b_{l'})\prod_{l=0}^{l'-1}b_l
$$
one has
\begin{eqnarray}
&&U(t_N,t_0)=\prod_{l=0}^{{N\over2}-1}U(t_{2l+2},t_{2l+1})U(t_{2l+1},t_{2l})\nonumber\\
&=&\prod_{l=0}^{{N\over2}-1}\tilde
U_{n_0+l}^A(t_{2l+2},t_{2l+1})U_{n_0+l}^A(t_{2l+1},t_{2l})
\nonumber\\ 
&+&\sum_{l=0}^{{N\over2}-1}\prod_{k=l+1}^{{N\over2}-1}U(t_{2k+2},t_{2k})
\left(U(t_{2l+2},t_{2l})-\tilde U_{n_0+l}^A(t_{2l+2},t_{2l+1})
U_{n_0+l}^A(t_{2l+1},t_{2l})\right)\times\nonumber\\
&\times &\prod_{k=0}^{l-1}\tilde
U_{n_0+k}^A(t_{2k+2},t_{2k+1})U_{n_0+k}^A(t_{2k+1},t_{2k})\nonumber\\
&=:&\prod_{l=0}^{{N\over2}-1}
\tilde U_{n_0+l}^A(t_{2l+2},t_{2l+1})U_{n_0+l}^A(t_{2l+1},t_{2l})+\RR_{\rm ad}(N)\label{U-UA}.
\end{eqnarray} 
The property $T^* U(t,s)T=U(t+1,s+1)$ leads to 
\begin{eqnarray*}
\RR_{\rm ad}(N)&:=&
\sum_{l=0}^{{N\over2}-1}U(t_N,t_{2l+2})(T^*)^l
\left( U(t_2,t_0))-\tilde U_{n_0+l}^A(t_{2},t_{1}) U_{n_0+l}^A(t_{1},t_{0})\right)T^l\times\\
&&\times\prod_{k=0}^{l-1}(T^*)^k\tilde
U_{n_0+k}^A(t_{2},t_{1})U_{n_0+k}^A(t_{1},t_{0})T^k.
\end{eqnarray*}
We pass now to the estimation of $\RR_{\rm ad}(N)$. Clearly do the presence of many unitarities
one gets at once from the adiabatic theorem:
\begin{eqnarray*}
\|\RR_{\rm ad}(N)\|&\le &
\sum_{l=0}^{{N\over2}-1}\left\| U(t_2,t_0)-\tilde U_{n_0+l}^A(t_{2},t_{1})
U_{n_0+l}^A(t_{1},t_{0})\right\|\\
&\le&\sum_{l=0}^{{N\over2}-1}
\left (\left\| U(t_2,t_1)-\tilde U_{n_0+l}^A(t_{2},t_{1})\right\|+\left\|
U(t_1,t_0)- U_{n_0+l}^A(t_{1},t_{0})\right\| \right )\\
&\le&\sum_{l=0}^{{N\over2}-1} 2C_Vb(n_0+l)\langle n_0+l\rangle ^{-1}.
\end{eqnarray*}
We concentrate in the following on the term
\begin{eqnarray*}
&&\prod_{l=0}^{\frac{N}{2}-1}
\tilde U_{n_0+l}^A(t_{2l+2},t_{2l+1})U_{n_0+l}^A(t_{2l+1},t_{2l})P_{2n_0-2}(t_0)\qquad
\\ &=&\sum_{j_{\frac{N}{2}}=1}^\infty P_{j_{\frac{N}{2}}}(t_{N})\times \\
&&\times\prod_{l=0}^{\frac{N}{2}-1}\left(
\tilde U_{n_0+l}^A(t_{2l+2},t_{2l+1})\sum_{{\tilde j}_l=1}^\infty
P_{{\tilde j}_l}(t_{2l+1})U_{n_0+l}^A(t_{2l+1},t_{2l})\sum_{j_l=1}^\infty 
P_{j_l}(t_{2l})\right)P_{2n_0-2}(t_0)\\
&=&\sum_{j_{\frac{N}{2}},
j_{\frac{N}{2}-1},{\tilde j}_{\frac{N}{2}-1},\ldots j_0,{\tilde j}_0}
P_{j_{\frac{N}{2}}}(t_{N})\times \\
&&\times\left(\prod_{l=0}^{\frac{N}{2}-1}\tilde
U_{n_0+l}^A(t_{2l+2},t_{2l+1})P_{\tilde j_l}(t_{2l+1})
U_{n_0+l}^A(t_{2l+1},t_{2l})P_{j_l}(t_{2l})\right)P_{2n_0-2}(t_0),
\end{eqnarray*}
where all the j's indices run for the moment from 1 to $\infty$. However since
 ${U_n^A}$ and ${\tilde U_n^A}$ leave invariant $\Ran Q_n$ and 
$\Ran {\tilde Q}_n$ and since for any $l\in\Int$
\begin{eqnarray*}
 Q_n(t_l)&=&P_{2n-2}(t_l)+P_{2n-1}(t_l)\\
 \tilde Q_n(t_l)&=&P_{2n-1}(t_l)+P_{2n}(t_l),
\end{eqnarray*}
we know that the only indices which
contribute are the ones of the set $J_{2n_0-2}$ where we define $J_n$ as follows:
$$
J_{n}:=\{(j_{\frac{N}{2}},\tilde j_{\frac{N}{2}-1},j_{\frac{N}{2}-1},\ldots ,\tilde j_0,j_0), 
j_0=2n_0-2,\,\forall l, \tilde
j_l=j_l
\ {\rm or}\ j_{l}+1\,;  j_{l+1}=\tilde j_l\ {\rm or}\ \tilde j_{l}+1 \}.
$$
Then ( we drop the times variables in $U^A$ and $\tilde {U}^A$ 
since their values are clear from the context)
$$
\left(\prod_{l=0}^{\frac{N}{2}-1} \tilde U_{n_0+l}^AU_{n_0+l}^A\right)P_{2n_0-2}=
\sum_{J_{2n_0-2}}P_{j_{\frac{N}{2}}}\prod_{l=0}^{\frac{N}{2}-1}\tilde  U_{n_0+l}^A
P_{\tilde j_l}U_{n_0+l}^AP_{j_l}.
$$
Clearly if we force $j_{\frac{N}{2}}$ to be equal to $2n_0+N-2$ by multiplying on the left by
$P_{2n_0+N-2}$ it remains only one "path" in this sum. Using (\ref{2.22})   
(\ref{2.38}), (\ref{calS}) and (\ref{caltildeS}) one may write
\begin{eqnarray*}
&&\langle \varphi_{2n_0+N-2}(t_N),\prod_{l=0}^{\frac{N}{2}-1} \tilde U_{n_0+l}^A(t_{2l+2},t_{2l+1})
U_{n_0+l}^A(t_{2l+1},t_{2l})\varphi_{2n_0-2}(t_0)\rangle\\
&&=\langle \varphi_{2n_0+N-2}(t_N),P_{2n_0+N-2}(t_N)\times \\
&&\times \prod_{l=0}^{\frac{N}{2}-1} \tilde
U_{n_0+l}^A(t_{2l+2},t_{2l+1})P_{2n_0+2l-1}(t_{2l+1})U_{n_0+l}^A(t_{2l+1},t_{2l})
P_{2n_0+2l-2}(t_{2l})\varphi_{2n_0-2}(t_0)\rangle \\
&&=\langle \varphi_{2n_0+N-2}(t_2),T^{\frac{N}{2}-1}P_{2n_0+N-2}(t_N)\times \\
&&\times \prod_{l=0}^{\frac{N}{2}-1}(T^*)^l\tilde U_{n_0+l}^A(t_{2},t_{1})
P_{2n_0+2l-1}(t_1)U_{n_0+l}^A(t_1,t_0)P_{2n_0+2l-2}(t_0)T^l\varphi_{2n_0-2}(t_0)\rangle\\
&&=\prod_{l=0}^{\frac{N}{2}-1}\tilde S_{2n_0+2l,2n_0+2l-1}^{n_0+l}(t_2,t_1)
S_{2n_0+2l-1,2n_0+2l-2}^{n_0+l}(t_1,t_0)\\
&&=:\prod_{l=0}^{\frac{N}{2}-1} e^{-i\tilde\theta_{n_0+l}(t_2,t_1)}\tilde\alpha_{2n_0+2l-1}(t_2,t_1)
e^{-i\theta_{n_0+l}(t_1,t_0)}\alpha_{2n_0+2l-2}(t_1,t_0)+\RR_{{\rm eff}}(N),
\end{eqnarray*}
where
\begin{eqnarray*}
|\RR_{{\rm eff}}(N)|&\le&\sum_{l=0}^{\frac{N}{2}-1}(2|\RR_{{\rm eff},n_0+l}|+|\RR_{{\rm eff},n_0+l}|^2)\\
&\le&\sum_{l=0}^\infty(2|\RR_{{\rm eff},n_0+l}|+|\RR_{{\rm eff},n_0+l}|^2)\\
&=&\frac{C_V\log^8\bra n_0\ket}{ \langle n_0 \rangle ^{\min\{1+8r, {1\over2}+5r, {1\over2}+r,1\}}}.
\end{eqnarray*}
Defining 
\begin{equation}
{\cal R}(N):={\cal R}_{{\rm eff}}(N)
+\langle \varphi_{2n_0+N-2},{\cal R}_{{\rm ad}}(N)\varphi_{2n_0-2} \rangle
\end{equation}
the lemma follows immediately from the estimates of ${\cal R}_{{\rm ad}}(N)$ and ${\cal R}_{{\rm eff}}(N)$.
\qed

\vskip 0.2cm
\noindent
{\em Proof of Theorem 6.1:} Let 
\begin{equation}\label{cal A}
{\cal A}(N):=\prod_{l=0}^{\frac{N}{2}-1}
e^{-i(\tilde\theta_{n_0+l}(t_2,t_1)+\theta_{n_0+l}(t_1,t_0))}
\tilde\alpha_{2n_0+2l-1}(t_2,t_1)\alpha_{2n_0+2l-2}(t_1,t_0).
\end{equation}
Then by Lemma 6.2 the propagating front reads as:
\begin{eqnarray}
{\cal P}(N)=|{\cal A}(N)|^2\cdot\left | 1+\frac{{\cal R}(N)}{{\cal A}(N)} \right |^2.
\end{eqnarray}
Replacing $\alpha_{2m-2}(t_1,t_0)$ and $\tilde\alpha_{2m-1}(t_2,t_1)$ for 
$m=n_0+k$ it follows that (we omit the time arguments for the simplicity of writing)  
\begin{eqnarray*}
&&|{\cal A}(N)|=\prod_{l=0}^{\frac{N}{2}-1}
|\tilde\alpha_{2n_0+2l-1}\alpha_{2n_0+2l-2}|\\
&&=\prod_{l=0}^{\frac{N}{2}-1}
\left ( \left (1-\frac {|{\hat V}(2n_0+2l-1)|^2}{n_0+l-\frac{1}{2}}
{\tilde c}_{2,n_0+l}\right )^2
+{\tilde\omega}_{1,n_0+l}^2
         \right )^{\frac{1}{2}}\times\\
&&\times\left ( \left (1-\frac{|{\hat V}(2(n_0+l-1))|^2}{n_0+l-1}c_{2,n_0+l} \right )^2+
\omega_{1,n_0+l}^2\right )^{\frac{1}{2}}\\
&&=\exp \left\lbrace \frac{1}{2}\sum_{l=0}^{\frac{N}{2}-1}
\log \left ( \left (1-\frac {|{\hat V}(2n_0+2l-1)|^2}{n_0+l-\frac{1}{2}}
{\tilde c}_{2,n_0+l}\right )^2
+{\tilde\omega}_{1,n_0+l}^2    \right )  
 \right \rbrace \times \\
&&\times  \exp \left\lbrace \frac{1}{2}\sum_{l=0}^{\frac{N}{2}-1}
\log \left ( \left (1-\frac{|{\hat V}(2(n_0+l-1))|^2}{n_0+l-1}c_{2,n_0+l} \right )^2+
\omega_{1,n_0+l}^2\right )     
\right \rbrace\\
&&=\exp \left\lbrace -\sum_{l=0}^{N-1}
\frac {|{\hat V}(2n_0+l-2)|^2}{n_0+\frac{l}{2}-1}c_{2,n_0+\frac{l}{2}}
 \right \rbrace\times
 \left (1+
{\cal O}\left ( \sum_{l=0}^{\infty}
\frac {1}{\langle n_0+l-\frac{1}{2}\rangle ^{1+2r}}     \right ) \right )
\end{eqnarray*}
where we used estimates of the following type (similar bounds are found for the sums 
containing $\omega_{1,n_0+l}^2$ and ${\tilde\omega}_{1,n_0+l}^2$) and 
the relation ${\tilde c}_{i,m}(t_2,t_1)=c_{i,m+\frac{1}{2}}(t_1,t_0)
$:
\begin{eqnarray*}
&&\sum_{l=0}^{\frac{N}{2}-1} \left |
\frac {|{\hat V}(2n_0+2l-1)|^2}{n_0+l-\frac{1}{2}}{\tilde c}_{2,n_0+l}  \right |
\leq \sum_{l=0}^{\infty}
\left |\frac {C_V}{\langle n_0+l-\frac{1}{2}\rangle ^{1+2r}}\right |
\end{eqnarray*}
Thus 
\begin{eqnarray*}
\lim_{N\to\infty}\left |\frac{{\cal R}(N)}{{\cal A}(N)}\right |=:|{\cal R}|\leq 
 \frac{ C_V\log ^8\langle n_0 \rangle }{\langle n_0
\rangle^{\min  \{1+8r,\frac{1}{2}+5r,\frac{1}{2}+r,1 \}} }\exp \left ( \sum_{l=0}^{\infty}
\frac {C_V}{\langle n_0+\frac{l}{2}-1\rangle ^{1+2r} } \right ).
\end{eqnarray*}
The proof of theorem is done.\,\qed
\par
\noindent
{\bf Remark 6.3:}
 Since as long as $r>0$ the series that appears in Eq.(\ref{propagf})  
both in the exponential and in the error term converges the propagating front 
does not vanish. This means that the electron is 
always accelerated and 
it escapes at infinity {\it in the energy space}, provided that it is initially
prepared in a band with sufficiently energy. 
 Thus we recovered rigorously the Ao's result.   
We end the paper with a result on the spectral properties of the Stark-Wannier
operator: 
\par
\vskip 0.2cm
\noindent
{\sl {\bf Corollary 6.4} For $r>0$ one has $\sigma_{{\rm cont}}(H^{SW})\neq\emptyset$.} 
\par
\nid
\vskip 0.2cm 
{\em Proof of Corollary 6.4:} Let $\phi\in {\cal H}_{{\rm pp}}(H^{SW})$ and $\Pi_N $ a family 
of bounded operators that goes strongly to zero when $N\to\infty $. Then it is known 
(see \cite{EV}) that  
$$
\lim_{N\to\infty}\sup_{\pm t>0}\|\Pi_N e^{-i t H^{SW}}\phi\|=0.
$$
In particular, for $t_N:=t_0+{N\over2}$
\begin{equation}\label{RAGEdiscret}
\lim_{N\to\infty}\|\Pi_N e^{-i (t_N-t_0) H^{SW}}\phi\|=0
\end{equation}
The strategy of the proof consists in writing $\Pi_N e^{-i (t_N-t_0) H^{SW}}\phi $ in the Fourier-Bloch 
representation and then to express the result in terms of the propagating front 
${\cal P}(N)$ which was computed in Thm. 6.1. Using (\ref{2.18}), writing (see (\ref{2.19}))
\begin{eqnarray*}
U(k,t_N,t_0)&=&U(0,t_N+k,t_0+k)=:U(t_N+k,t_0+k)\\
&=&U(t_N+k,t_N)U(t_N,t_0)U(t_0,t_0+k)
\end{eqnarray*}  
and chosing
\begin{equation}\label{chose} 
\phi:= e^{i t_0 X} S^*  \left(\int_{[0,1)}^\oplus dk\,
U(t_0,t_0+k)\varphi_{2n_0-2}(t_0)\right)
\end{equation}
one has
\begin{eqnarray*}
&&\|\Pi_N e^{-i (t_N-t_0) H^{SW}}\phi\|\\
&=&\| S e^{-it_N X}\Pi_Ne^{it_N X}S^*
\left(\int_{[0,1)}^\oplus dk\,U(t_N+k,t_N)U(t_N,t_0)\varphi_{2n_0-2}(t_0)\right)\|.
\end{eqnarray*}
Now we chose $\Pi_N$ such that:
\begin{equation}\label{PiN}
 S e^{-it_N X}\Pi_Ne^{it_N X}S^*
\left(\int_{[0,1)}^\oplus dk\,U(t_N+k,t_N)\right)=
\left(\int_{[0,1)}^\oplus dk\, P_{2n_0-2+N}(t_N)\right).
\end{equation}
Let us show that $\Pi_N$ goes strongly to zero as $N$ goes to infinity. Using the identity 
\begin{equation}
Se^{-in X}S^*=\id\otimes T^{-n},\qquad \forall n\in\Int
\end{equation}
it turns out that $\Pi_N$ is given by 
\begin{eqnarray*}
\Pi_N&:=& e^{it_N X} S^* \left(\int_{[0,1)}^\oplus
dk\,P_{2n_0-2+N}(t_N)U(t_N,t_N+k)\right)Se^{-it_N X}\\
&=&
e^{it_0 X} S^*\left(\int_{[0,1)}^\oplus
dk\, T^{\frac{N}{2}}P_{2n_0-2+N}(t_N)U(t_N,t_N+k)T^{-\frac{N}{2}}\right )Se^{-it_0 X}\\
&=&e^{it_0 X}
S^* \left(\int_{[0,1)}^\oplus
dk\, P_{2n_0-2+N}(t_0)U(t_0,t_0+k)\right)Se^{-it_0 X}.
\end{eqnarray*}
Then clearly
$$
s-\lim_{N\to\infty}\Pi_N=0.
$$
Finally observe that from (\ref{chose}) and (\ref{PiN})
\begin{eqnarray*}
\|\Pi_N e^{-i (t_N-t_0)
H^{SW}}\phi\|^2=\|P_{2n_0-2+N}(t_N)U(t_N,t_0)\varphi_{2n_0-2}(t_0)\|^2={\cal P}(N).
\end{eqnarray*}
However, from Thm.6.1 we know that for $n_0$ sufficiently large and $r>0$ the propagating front ${\cal P}(N)$
does not vanish when $N\to\infty$ so $\phi$ must have a part in ${\cal H}_{{\rm cont}}(H^{SW})$.\qed 

When applied to the driven quantum ring problem the existence of a propagating front gives at once:

\vskip 0.2cm
\noindent
{\sl {\bf Corollary 6.5} Let $r>0$ and $\varphi(t_0=-\frac{1}{4})=\varphi_{2n_0-2}(t_0)$,
$\varphi(t_N=t_0+\frac{N}{2})=U(t_N,t_0)\varphi(t_0)$. Then

\begin{equation}
\langle \varphi(t_N),H(t)\varphi(t_N)\rangle\geq 
\left (1+{\cal O}_r(\frac{1}{\langle n_0 \rangle ^r}) \right )\cdot t^2.
\end{equation}

}

We end up this section of applications by showing that there is no localization
in momentum space for appropriate initial conditions.
\vsth
\nid{\bf Corollary 6.6}
{\sl\ Let $F_D(I)$ be the spectral projection of $D:=-i\partial_x$ on
the interval$I:=[n_0,n_0+1)$. Then for all $V$ such that $r>0$ one has$$
\lim_{n_0\to\infty}\lim_{N\to\infty}\|F_D(I+t_N)e^{-i(t_n-t_0)H^{SW}}F_D(I+t_0)\|=1.$$ }
\vsth\nid
{\em Proof}: One can check that $SDS^\star$ is simply the multiplication by $k+n$;
thus $SF_D(I)S^\star=1\otimes P_{n_0,0}$. One has successively:
\begin{eqnarray*}&&\|F_D(I+t_N)e^{-i(t_N-t_0)H^{SW}}F_D(I+t_0)\|
\\&=&\|F_D(I+t_N)e^{it_NX}S^\star\int_{[0,1)}^\oplus dkU(t_N+k,t_0+k) Se^{-it_0X}F_D(I+t_0)\|
\quad\mbox{(see (\ref{2.18}))}\\&=&\|Se^{-it_NX}F_D(I+t_N)e^{it_NX}S^\star\int_{[0,1)}^
\oplus dkU(t_N+k,t_0+k) Se^{-it_0X}F_D(I+t_0)e^{it_0X}S^\star\|
\\&=&\|SF_D(I)S^\star\int_{[0,1)}^\oplus dkU(t_N+k,t_0+k) SF_D(I)S^\star\|
\\&=&\|1\otimes P_{n_0,0}\int_{[0,1)}^\oplus dkU(t_N+k,t_0+k) 1
\otimes P_{n_0,0}\|=\sup_{k\in[0,1)}\| P_{n_0,0}U(t_N+k,t_0+k) P_{n_0,0} \|
\\&\ge& \|P_{n_0,0} U(t_N,t_0)P_{n_0,0}\|=\|P_{2n_0+N}^0(t_N)U(t_N,t_0)P_{2n_0+N}^0(t_0)\|
\quad\mbox{(see (\ref{labelling1}))}\\
&=&\|P_{2n_0+N}(t_N)U(t_N,t_0)P_{2n_0+N}(t_0)\|
+\OO_r(b(n_0+1))\qquad\mbox{(see Corollary 2.6)}
\\&=&{\cal P}(N+1)+\OO_r(b(n_0+1)).
\end{eqnarray*}
Using theorem 6.1, the rest is now clear.\qed

\setcounter{equation}{0}
\section{Appendices}
\numberwithin{equation}{subsection}
\subsection{The key estimate }

 \noindent
{\em Proof of Lemma 2.3: } 
\vskip 0.2cm
In the following we shall use the notation $\alpha:=\max\{0,-r\}$.
We start by estimating $\sup_{t\in I_0}\sup_{z\in\Gamma_m}\|K((d_m+iy),t)\|$:

\begin{eqnarray}\nonumber
&&\sup_{t\in I_0}\sup_{z\in\Gamma_m}\|K((d_m+iy),t)\|_{HS}\leq
\|V\|_r\left(\sqrt{5}\over2\right)^{\alpha }\sup_{t\in I_0}\sum_{n\in\Int}
{\bra n\ket^{2\alpha}\over|(n+t)^2-d_m-iy|}\\\nonumber
&\le&\|V\|_r\left(\sqrt{5}\over2\right)^{\alpha }\sup_{t\in I_0}\sum_{n\in\Int}
{\bra n\ket^{2\alpha}\over |(n+t)^2-d_m|}
\leq 4\|V\|_r\left(\sqrt{5}\over2\right)^{\alpha }\sum_{n\in\Int}
{\bra n\ket^{2\alpha}\over |n^2-d_m|}\\\label{suma1}
&=&\|V\|_r\left(\sqrt{5}\over2\right)^{\alpha }
\left\lbrace \frac{1}{d_m}+2\sum_{n\geq 1}{\bra n\ket^{2\alpha}\over |n^2-d_m|}
     \right\rbrace
\end{eqnarray}
where the time dependence of $K((d_m+iy),t)$ was eliminated by using
 the inequality (proved bellow )
\begin{equation}\label{elim}
\sup_{t\in I_0}|(n+t)^2-d_m|\geq \frac{1}{4}|n^2-d_m|.
\end{equation}

The next step is to estimate the sum from the RHS of (\ref{suma1}). Let us
assume that $m>1$ (the particular case $m=1$ will be discussed separately).
Then approximating the sum with integrals one has 
\begin{eqnarray}\nonumber
 \sum_{n\ge1}\frac{\langle m\rangle ^{2\alpha}}{|n^2-d_m|}&\le&
 \int_1^{m-1}\frac {\langle
x\rangle^{2\alpha}}{ d_m-x^2}dx+
\frac{\langle m-1\rangle^{2\alpha}}{\frac{1}{2}(2m-\frac{3}{2})}
+\frac{\langle m\rangle^{2\alpha}}{m-\frac{1}{4}}+
\int_m^{\infty}\frac{ \langle x\rangle^{2\alpha}}{x^2-d_m}\\\label{suma2}
&=:&S_1+S_2+S_3+S_4.
\end{eqnarray}

Since $\langle x\rangle $ is an increasing function we can estimate the term $S_1$
as follows
\begin{eqnarray}\nonumber
S_1&\leq&\langle m-1\rangle^{2\alpha}\int_1^{m-1}\frac {dx}{ d_m-x^2}\\\nonumber
&=&\frac{\langle m-1\rangle^{2\alpha}}{2\sqrt {d_m}}
\log \left (\frac{\sqrt {d_m}+m-1}{\sqrt {d_m}-(m-1)}\cdot\frac{\sqrt {d_m}-1}
{\sqrt {d_m}+1} \right )\\\label{S1}
&\leq&\frac{\langle m\rangle ^{2\alpha}}{2\sqrt {d_m}}\log \left (\frac{m+\sqrt {d_m}}{m-\sqrt {d_m}}
\right ).
\end{eqnarray}
To obtain the last line we used the following estimate (valid for $m\geq 2$):
\begin{eqnarray*}
&&\frac{\sqrt {d_m}+m-1}{\sqrt {d_m}-(m-1)}\cdot\frac{\sqrt {d_m}-1}
{\sqrt {d_m}+1}=\frac{2m-\frac{3}{2}  }{\frac{1}{2}}\cdot\frac{m-\frac{3}{2} }{m+\frac{1}{2} }\\
&&\leq\frac{2m-\frac{1}{2}  }{\frac{1}{2}}=\frac{m+\sqrt {d_m}}{m-\sqrt {d_m}}.
\end{eqnarray*}
The next two terms are easier:
\begin{eqnarray}\label{S2}
S_2&\leq&\frac{12}{5\log 7}\cdot\frac{\langle m \rangle ^{2\alpha}}{2\sqrt {d_m}}
\cdot\log \left (\frac{m+\sqrt {d_m}}{m-\sqrt {d_m}}\right )\\\label{S3}
S_3&\leq&\frac{12}{7\log 7}\cdot\frac{\langle m \rangle ^{2\alpha}}{2\sqrt {d_m}}
\cdot\log \left (\frac{m+\sqrt {d_m}}{m-\sqrt {d_m}}\right ).
\end{eqnarray}
We give some details only for $S_2$. First we write:
\begin{eqnarray*}
S_2=\frac{\langle m-1\rangle^{2\alpha}}{\frac{1}{2}(2m-\frac{3}{2})}
\leq \frac{\langle m\rangle^{2\alpha}}{\frac{1}{2}(2m-\frac{3}{2})}
\cdot\frac{\log \left (\frac{m+\sqrt {d_m}}{m-\sqrt {d_m}}\right )}{2\sqrt {d_m}}
\cdot \frac{2\sqrt {d_m}  }{ \log \left (\frac{m+\sqrt {d_m}}{m-\sqrt {d_m}}\right )}.  
\end{eqnarray*}
Then use
\begin{eqnarray*}
&&\frac{2\sqrt {d_m}  }{\frac{1}{2}(2m-\frac{3}{2}) }
=\frac{4(m-\frac{1}{2})}{2m-\frac{3}{2}}=2\cdot\frac{m-\frac{1}{2} }{m-\frac{3}{4}}
\leq \frac{12}{5}
\end{eqnarray*}
and 
\begin{eqnarray*}
\frac{1}{\log \left (\frac{m+\sqrt {d_m}}{m-\sqrt {d_m}}\right )}
=\frac{1}{\log \left (\frac{2m-\frac{1}{2}}{\frac{1}{2}} \right )}\leq
\frac{1}{\log 7}.
\end{eqnarray*}
To estimate $S_4$ we make an integration by parts:
\begin{eqnarray}\nonumber
S_4&=&\int_m^{\infty}\frac{(1+x^2)^{\alpha }}{x^2-d_m}\\\nonumber
&=&\left [\frac{(1+x^2)^{\alpha }}{2\sqrt {d_m}}
\log \left (\frac{x-\sqrt {d_m}}{x+\sqrt {d_m}}\right )\right ]_m^{\infty}
-\frac{\alpha}{\sqrt {d_m}}\int_m^{\infty}dx x{(1+x^2)^{\alpha -1}}
\log \left (\frac{x-\sqrt {d_m}}{x+\sqrt {d_m}}\right )\\\nonumber
&\leq&\frac{\langle m \rangle ^{2\alpha}}{2\sqrt {d_m}}
\log \left (\frac{m+\sqrt {d_m}}{m-\sqrt {d_m}}\right )
+\frac{\alpha}{\sqrt {d_m}}\left (\sup_{x\geq m} x
\log \left (\frac{x+\sqrt {d_m}}{x-\sqrt {d_m}}\right )   \right )
\int_m^{\infty}x^{2(\alpha-1)}dx.
\end{eqnarray}
The supremum above is computed by simply noticing that the function
$f(x):=x\log (\frac{x+a}{x-a})$ is decreasing on $(a,\infty )$. Indeed, its first two derivatives
are given by
\begin{eqnarray*}
f'(x)&=&\log \left (\frac{x+a}{x-a}\right )+\frac{2xa}{a^2-x^2}\to 0\qquad {\rm if}\,\, x\to\infty\\
f''(x)&=&\frac{4a^3}{(a^2-x^2)^2}.
\end{eqnarray*}
Hence
\begin{eqnarray}\nonumber
S_4&\leq&\frac{\langle m \rangle ^{2\alpha}}{2\sqrt {d_m}}
\log \left (\frac{m+\sqrt {d_m}}{m-\sqrt {d_m}}\right )
+\frac{\alpha}{1-2\alpha}\frac{\langle m \rangle ^{2\alpha}}{\sqrt {d_m}}
\log \left (\frac{m+\sqrt {d_m}}{m-\sqrt {d_m}}\right )
\\\label{S4}
&=&\frac{1}{2(1-2\alpha)}\cdot\frac{\langle m \rangle ^{2\alpha}}{2\sqrt {d_m}}
\cdot\log \left (\frac{m+\sqrt {d_m}}{m-\sqrt {d_m}}\right ).
\end{eqnarray}
Now taking into account that
\begin{equation}\label{1dm}
\frac{1}{d_m}\leq
\frac{4}{3\log 7}\frac{\langle m \rangle ^{2\alpha}}{2\sqrt {d_m}}
\log \left (\frac{m+\sqrt {d_m}}{m-\sqrt {d_m}}\right ),
\end{equation}
and that $2\sqrt {d_m}\geq \frac{3}{\sqrt 5}\langle m \rangle$
we collect all the estimates (\ref{S1}),(\ref{S2}),(\ref{S3}),(\ref{S4}) and plug
them into (\ref{suma1}):
\begin{eqnarray}\nonumber
\sup_{t\in I_0}\sup_{z\in\Gamma_m}\|K((d_m+iy),t)\|
&\leq&\|V\|_r\frac{4\sqrt 5}{3}\left(\sqrt{5}\over2\right)^{\alpha }\cdot
\frac{9-6\alpha}{1-2\alpha}
\frac{\log (4m-1)}{\langle m \rangle ^{1-2\alpha}}\\\label{ka1}
&:=&C_V\frac{\log (4m-1)}{\langle m \rangle ^{1+\min\{0,2r\}}}.
\end{eqnarray}
 Note that the constant $C_V$ depends on $r$
and has a pathological behaviour at $r=-\frac{1}{2}$.
We treat now the case $m=1$. This term contains actually only the terms
of the type $S_3$ and $S_4$, so that:
\begin{eqnarray}\nonumber
\sup_{t\in I_0}\sup_{z\in\Gamma_1}\|K((d_1+iy),t)\|
\leq\|V\|_r\left(\sqrt{5}\over2\right)^{\alpha }
\left\lbrace \frac{1}{d_1}+2\sum_{n\geq 1}{\bra 1\ket^{2\alpha}\over |n^2-d_1|}
     \right\rbrace
\end{eqnarray}
and
\begin{eqnarray}\nonumber
\sum_{n\geq 1}^{\infty}\frac{\langle n\rangle ^{2\alpha}}{|n^2-\frac{1}{4}|}
\leq\frac{4\langle 1\rangle ^{2\alpha}}{3}+\int_2^{\infty}
\frac{\langle x\rangle ^{2\alpha}}{x^2-\frac{1}{4}}.
\end{eqnarray}
This leads to
\begin{eqnarray}\nonumber
\sup_{t\in I_0}\sup_{z\in\Gamma_1}\|K((d_1+iy),t)\|
&\leq&\|V\|_r\left(\sqrt{5}\over2\right)^{\alpha }
\left (4+\frac{8\langle 1\rangle ^{2\alpha}}{3}+
\frac{\langle 1\rangle ^{2\alpha}}{1-2\alpha}\log 3 \right )\\\label{m=1}
&\leq&C_V
\frac{\log 3}{\langle 1\rangle ^{1+\min\{0,2r\}}}.
\end{eqnarray}
Inspecting (\ref{ka1}) and (\ref{m=1}) it is clear that one has for any
$m\geq 1$
\begin{equation}\label{ka2}
\sup_{t\in I_0}\sup_{z\in\Gamma_m}\|K((d_m+iy),t)\|
\leq C_V\frac{\log (4m-1)}{\langle m \rangle ^{1+\min\{0,2r\}}}.
\end{equation}
 From this estimate one can identify $m^*$ such that for any $m>m^*$ we have
$\sup_{t\in I_0}\sup_{z\in\Gamma_m}\|K((d_m+iy),t)\|\leq 1$.

Now we turn to the estimation of
$ \sup_{t\in I_0}\int_{\Gamma_m}\|R_0(z,t)\|\cdot\|K(z,t)\|dy$.
The resolvent is easily estimated as
\begin{equation}
\|R_0(z,t) \|\leq\frac{1}{\sqrt{A^2+y^2}  },
\end{equation}
where we introduced the notation $A:=(m-\frac{1}{2})^2-(m-\frac{3}{4})^2$. Since
$\sup_{t\in I_0}\|K(z,t)\|\leq 4 \|K(z,0)\|_{HS}$ we have
\begin{eqnarray}\nonumber
&&\sup_{t\in I_0}\int_{\Gamma_m}\|R_0(z,t)\|\cdot\|K(z,t)\|dy\\\nonumber
&\leq & 4\|V\|_r\left(\sqrt{5}\over2\right)^{\alpha }
\int_{-\infty}^{\infty}dy\frac{1}{\sqrt{A^2+y^2} }
\left (\sum_{n_1\in\Int}\frac{\langle n_1 \rangle ^{2\alpha}}{\sqrt{B_{n_1}^2+y^2}}
\cdot\sum_{n_2\in\Int}\frac{\langle n_2 \rangle ^{2\alpha}}
{\sqrt{B_{n_2}^2+y^2}}\right )^{\frac{1}{2}}\\\nonumber
&\leq & 4\|V\|_r\left(\sqrt{5}\over2\right)^{\alpha }
\left (\int_{-\infty}^{\infty}dy \sum_{n_1\in\Int}
\frac{\langle n_1 \rangle ^{2\alpha}}{ \sqrt{A^2+y^2}\sqrt{B_{n_1}^2+y^2}}\right )^{\frac{1}{2}}\\\nonumber
&\cdot & \left (\int_{-\infty}^{\infty}dy \sum_{n_2\in\Int}
\frac{\langle n_2 \rangle ^{2\alpha}}{ \sqrt{A^2+y^2}\sqrt{B_{n_2}^2+y^2}}   \right )^{\frac{1}{2}}
\qquad {\rm by\, Cauchy-Schwartz}
\\\nonumber
&=& 4\|V\|_r\left(\sqrt{5}\over2\right)^{\alpha }
\sum_{n\in\Int}\langle n \rangle ^{2\alpha} \int_{-\infty}^{\infty}
\frac{dy}{ \sqrt{A^2+y^2}\sqrt{B_{n}^2+y^2}}\\\nonumber
&:=&4\|V\|_r\left(\sqrt{5}\over2\right)^{\alpha }
\sum_{n\in\Int}\langle n \rangle ^{2\alpha}I(A,B_n).
\end{eqnarray}
In the above calculation we used as well the notation:
\begin{equation}
B_n:=n^2-d_m.
\end{equation}
The integral $I(A,B_n)$ is estimated as follows (in our case $B_n>A$ always)
\begin{eqnarray*}
  I(A,B_n)
  &\leq &\frac{1}{B_n^{1-\delta }}\int_{-\infty}^{\infty}dy\frac{1}
 {\sqrt{A^2+y^2}\cdot(A^2+y^2)^{\frac{\delta}{2} }   }\\
 &=& \frac{1}{A^{\delta }B_n^{1-\delta }}\int_{-\infty}^{\infty}dp\frac{1}
{(1+p^2)^{\frac{1+\delta }{2} }}\\
 &=& \frac{1}{A^{\delta } B_n^{1-\delta }}\cdot \frac{ \sqrt{\pi}\Gamma (\delta /2) }
 {\Gamma ((1+\delta) /2)}=\frac{C_{\delta }}{A^{\delta } B_n^{1-\delta}},
\end{eqnarray*}
 where we made the substitution $p=y/A$. On the other hand $B_n$ obeys the
following estimate
\begin{equation}
B_n=|n^2-d_m|\le\max\{n^2,d_m\}\le\max\{\bra n\ket^2,\bra d_m\ket^2\}\le\bra
n\ket^2\bra d_m\ket^2
\end{equation}
from where
\begin{equation}
 I(A,B_n)\leq C_{\delta }\frac{ \bra n\ket^{2\delta} \bra d_m\ket^{2\delta}}{A^{\delta }B_n}.
\end{equation}
Using the fact that the function $g(m):=\frac{\bra d_m\ket^2}{A\bra n\ket}$ is decreasing
so that $\max_{m\in\Int}g(m)=g(0)=4$ one has
$\frac{\bra d_m\ket^{2\delta}}{A^{\delta }}\leq 4\bra m \ket^{2\delta }$. Thus
\begin{eqnarray}\nonumber
&&\sup_{t\in I_0}\int_{\Gamma_m}\|R_0(z,t)\|\cdot\|K(z,t)\|dy
\leq 16\|V\|_r\left(\sqrt{5}\over2\right)^{\alpha }C_{\delta } \bra m \ket^{2\delta }
\sum_{n\in\Int}\frac{\langle n \rangle ^{2(\alpha +\delta)} }{|n^2-d_m|}\\
&&\leq C_VC_{\delta } \frac{\langle m \rangle ^{4\delta}
\log (4m-1)}{\langle m \rangle ^{1+\min\{0,2r\}}}\qquad 0<\delta<
\frac{1}{4}-\frac{\alpha}{2}.
\end{eqnarray}
The last condition on $\delta$ assures that the RHS vanishes in the limit
$m\to\infty$. We further optimise the result by using the inequality
\begin{equation}
C(\delta)\leq\frac{3}{\delta}
\end{equation}
and taking
\begin{equation}
\delta =\frac{1}{4\log \bra m\ket}<\frac{1}{4}-\frac{\alpha}{2},
\end{equation}
from where it follows that $\bra m\ket\ge e^{\frac{1}{(1-2\alpha)}}$ and
\begin{equation}
\delta^{-1}\bra m\ket ^{4\delta}=4e\log \bra m\ket
\end{equation}
Then
\begin{equation}\label{4delta}
\sup_{t\in I_0}\int_{\Gamma_m}\|R_0(z,t)\|\cdot\|K(z,t)\|dy
\leq C_V\frac{\log ^2 (4m-1)}{\bra m\ket ^{\min \{1,1+2r\}}   }
\end{equation}
\\
\nid 
The estimates (\ref{ka2}) and (\ref{4delta}) lead to
\begin{equation}\label{l3.2}
\sup_{t\in I_0}\left (\sup_{z\in\Gamma_m}\|K(z,t)\|
+\int_{\Gamma_m}\|R_0(z,t)\|\cdot\|K(z,t)\|dy  \right )
\leq C_V\frac{\log ^2\langle 4m-1 \rangle}
{\langle m \rangle ^{\min \{1,1+2r \}}}.
\end{equation}
In what concerns the estimates on ${\tilde\Gamma}_m$ one has to follow the same
steps as above, the result being that $m$ is pushed to $m+\frac{1}{2}$
Since
$\langle m+\frac{1}{2} \rangle\leq \sqrt 2 \langle m \rangle$ and
$\log (4m+1)\leq 2\log (4m-1)$ one can work only with
$b(m)$.
To prove (\ref{2.32'}) and (\ref{aaa}) one has to use the first resolvent identity:
\begin{eqnarray*}
\|K(z,t)\|&\leq& \|K(d_m,t)\|\cdot \left \| \frac{R_0(z,t)}{R_0(d_m,t)} \right \|\\
&\leq&\|K(d_m,t)\| \left (1+\frac{|z-d_m|}{\inf_{z\in\gamma_m}{\rm dist}
(z,\sigma (H_0(t)) )  }       \right ).
\end{eqnarray*}
Since
\begin{eqnarray*}
\sup_{z\in\gamma_n}|z-d_m|=\sqrt 5 (m-1)\\
\inf_{t\in I_0}{\inf_{z\in\gamma_m}{\rm dist}(z,\sigma (H_0(t)) )  }
\leq \frac{1}{4}(2m-\frac{11}{4})
\end{eqnarray*}
from where (\ref{aaa}) follows.
Finally let us prove (\ref{elim})
 \begin{eqnarray*}
\inf_{t\in I_0}\inf_{n\in\Int}\left|(n+t)^2-d_m\over n^2-d_m\right|
&=&\inf_{t\in I_0}\inf_{n\in\Int}\left|(n+t)-\sqrt{d_m}\over n-\sqrt{d_m}\right|
\left|(n+t)+\sqrt{d_m}\over
n+\sqrt{d_m}\right|\\
&\ge&\inf_{t\in I_0}\inf_{n\in\Int}\left|(n+t)-\sqrt{d_m}\over n-\sqrt{d_m}\right|
\inf_{t\in I_0}\inf_{n\in\Int}\left|(n+t)+\sqrt{d_m}\over
n+\sqrt{d_m}\right|\\
&=&\inf_{t\in I_0}\inf_{n\in\Int}\left|1+{t\over n-\sqrt{d_m}}\right|
\inf_{t\in I_0}\inf_{n\in\Int}\left|1+{t\over n+\sqrt{d_m}}\right|\\
&\ge&\left (1-\sup_{t\in I_0}\sup_{n\in\Int}\left|{t\over n-\sqrt{d_m}}\right| \right)
\left(1-\sup_{t\in I_0}\sup_{n\in\Int}\left|{t\over n+\sqrt{d_m}}\right|\right )\\
&\ge&\left(1-\sup_{n\in\Int}\left|{{1\over4}\over n-\sqrt{d_m}}\right|\right )
\left(1-\sup_{n\in\Int}\left|{{1\over4}\over n+\sqrt{d_m}}\right|\right)\\
&=&\left (1-{1\over 4\dist(\sqrt{d_m},\Int)} \right )^2={1\over4}.
\end{eqnarray*}

The proof of Lemma 2.3 is finished.\qed 
 
\setcounter{equation}{0}

\subsection{The Sz-Nag\"y tranformation }
                 
For convenience we give here the well known \cite{K} Sz-Nag\"y lemma 
on unitary equivalence of orthogonal projections.
\vskip 0.2cm
\noindent
{\sl {\bf Lemma 7.2.1: } Let $P,Q$ be orthogonal projections in a Hilbert space ${\cal H} $ satisfying
\begin{equation}
\|P-Q\|<1
\end{equation} 
If $U$ is defined by 
\begin{equation}
U=(1-(P-Q)^2  )^{-\frac{1}{2}}(PQ+(1-P)(1-Q))
\end{equation}
then 
\begin{equation}
 U^*=U^{-1}
\end{equation}
and 
\begin{equation}
P=UQU^*
\end{equation}
Notice also that
\begin{equation}\label{A.5}
(1-(P-Q)^2)^{-\frac{1}{2}}=
1+\sum_{j\geq 1}\frac{(2j-1)!!}{j!2^j}(P-Q)^{2j}.
\end{equation}}

\setcounter{equation}{0}
\subsection{The remainder of order $p$ for the Dyson series }

In this appendix we review a method to estimate the remainder of order $p$ for the Dyson 
series of ${\widehat\Omega}_{P,m}$ encountered in Section 5. 
Let $\EE$ be the space of $C^\infty$ operator-valued functions on  $I_0$ 
taking values in ${\cal B}(\Com ^2)$ . Thus $\forall X\in \EE,\forall l\in\Nat\quad$
\begin{equation}
 \|X\|_\infty:=\sup_{t\in I_0}\|\partial_t^lX(t)\|<\infty,
\end{equation} 
since $I_0$ is compact and $t\to \partial_t^l X(t)$ is continuous. 
We define the  operator $K:\EE\to\EE$ by the following relation
\begin{equation}
(KX)(t):=-i\int_{t_0}^t{\widehat V}_P(s) X(s)ds,\quad
{\widehat V}_P(t):=\left (\begin{array}{cc}0&b(t)\\
                      \overline{b(t)}&0\end{array}\right ),
\end{equation} 
where $b(t):={\hat V}(2(m-1))e^{2i(m-1)(t^2-t_0^2)}$. 
The Dyson series with remainder of order $p$ reads as
\begin{equation}  
{\widehat\Omega}^0_{P,m}(t_1,t_0)=\sum_{k=0}^{m-1}K^k(t)\id +K^m(t){\widehat\Omega}^0_{P,n_0}(t_1,t_0).
\end{equation} 
Here $K^k$ denotes  $K\circ K\circ \ldots \circ K$ k times. 
We start by proving a technical lemma.
\vsth
{\sl {\bf Lemma 7.3.1} Let $a:=m-1$. Then for all $X\in \EE $ it holds 
\begin{equation} 
\|KX\|_\infty\le
{|b|\over\sqrt{a}}2\sqrt{6}\|X\|_\infty+|b|{\log({\sqrt{a}\over {2\sqrt{6}}}\vee
1)\over a}\|\dot X\|_\infty ,
\end{equation}
where we used the notation $m\vee n:=\max \{m,n\}$ }  
\vsth
\nid 
{\em Proof:} We shall use the canonical basis in 
$\BB(\C^2)$ which reads as follows
\begin{equation} 
\Pi_{1,1}:=\left (\begin{array}{cc}1&0\\
                 0&0\end{array}\right )\quad
\Pi_{2,2}:=\left (\begin{array}{cc}0&0\\
                 0&1\end{array}\right )\quad
\Pi_{1,2}:=\left (\begin{array}{cc}0&1\\
                 0&0\end{array}\right )\quad
\Pi_{2,1}:=\left (\begin{array}{cc}0&0\\
                 1&0\end{array}\right )
\end{equation}
Also, let $\chi_-$, $\chi_0$ et $\chi_+$ be the characteristic functions 
corresponding to the intervals $[-{1\over4},-\eps]$, $[-\eps,\eps]$ and
$[\eps,{1\over4}]$ where $0<\eps<{1\over4}$.
Then we have
\begin{equation}   
KX=K\chi_-X+K\chi_0X+K\chi_+X.
\end{equation} 
One gets immediately
\begin{equation} 
\|K\chi_0X\|_\infty\le 2\eps |b|\|X\|_\infty.
\end{equation}
For $K\chi_\pm X$ we use the identity 
$K\chi_\pm X=K\chi_\pm\Pi_{1,1}X+K\chi_\pm\Pi_{2,2}X$.
We shall treat in detail only the term $K\chi_-\Pi_{2,2}X$, 
the estimates for the remaining parts being completely similar.
Then with the notation 
$m\wedge n:=\min\{m,n\}$ we have 
\begin{eqnarray*}
(K\chi_-\Pi_{2,2}X)(t)
&=&-i\int_{t_0}^{t\wedge -\eps} b(s)\Pi_{1,2}X(s)ds\\
&=&-\int_{t_0}^{t\wedge -\eps} 4ias\, b(s){\Pi_{1,2}X(s)\over 4 a s}ds\\
&=&-\left[b(s)\Pi_{1,2}X(s)\over4 a s\right]_{t_0}^{t\wedge -\eps} +
\int_{t_0}^{t\wedge -\eps}
b(s)\partial_s\left(\Pi_{1,2}X(s)\over 4 as\right)ds,
\end{eqnarray*}
since $\dot b(s)=4ias b(s)$.
The first term is estimated as follows
\begin{equation}
\left\|\left[b(s){\Pi_{1,2}X(s)\over
4as}\right]_{t_0}^{t\wedge -\eps}\right\|\le 
{|b|\over 2a\eps}\|\Pi_{1,2}X\|_\infty.
\end{equation}
For the second term
\begin{equation}
\int_{t_0}^{t\wedge -\eps}
b(s)\partial_s\left(\Pi_{1,2}X(s)\over 4 as\right)ds=
\int_{t_0}^{t\wedge
-\eps} b(s)\left(-{\Pi_{1,2}X(s)\over 4 as^2}+{\Pi_{1,2}\dot X(s)\over 4
as}\right)ds,
\end{equation}
and consequently
\begin{eqnarray*}
\left\|\int_{t_0}^{t\wedge -\eps}
b(s)\partial_s\left(\Pi_{1,2}X(s)\over 4 as\right)ds\right\|
&\le& \|\Pi_{1,2}X\|_\infty
\int_{t_0}^{-\eps} {|b|\over 4a s^2}ds+\|\Pi_{1,2}\dot
X\|_\infty\int_{t_0}^{-\eps}{|b|\over 4a |s|}ds\\ 
&\le&{|b|\over 4a\eps}  \|\Pi_{1,2}X\|_\infty-{|b|\log(4\eps)\over 4a} 
\|\Pi_{1,2}\dot X\|_\infty.
\end{eqnarray*}
Finally we find 
\begin{equation}
\|KX\|_\infty\le |b|\left( (2\eps+{1\over a\eps})
\|X\|_\infty-{\log(4\eps)\over
a}\|\dot X\|_\infty\right)
\end{equation} 
and chosing  $\eps=\sqrt 3(2a)^{-{1\over2}}\wedge {1\over 4}$ 
we arrive at the announced result.\qed
\\
Using Lemma 7.3.1 and the identity $\frac{d}{dt}(K^{p-1}X)(t)=\widehat V(t)(K^{p-2}X)(t)$ 
one can get the general result:
\begin{eqnarray}\nonumber
{\rm if}\ p\ge2,\quad\|K^pX\|_\infty
&\le&
{|b|\over\sqrt{a}}c\|K^{p-1}X\|_\infty+|b|{\log{\sqrt{a}\over c}\over
a}\|\widehat VK^{p-2}X\|_\infty\\\label{recur}
&\le&A\|K^{p-1}X\|_\infty+B\|K^{p-2}X\|_\infty ,
\end{eqnarray}
with the notations
\begin{eqnarray}\nonumber
c=2\sqrt 6, A:=\frac{|b|}{\sqrt a}c, B=|b|^2\frac{\log \frac{\sqrt a}{c}\vee 1 }{a}. 
\end{eqnarray}
By looking at the expansion of $\Omega^0_{P,m}(t_1,t_0)$ it is clear that we have to solve  
(\ref{recur}) in the case $X=\Omega^0_{P,m}$. However, in order to estimate as well 
$\widehat\Omega_{P,m}^{0,(p)}$ the case $X=1$ would be also usual. 
We consider then the following two sequences
for $\forall p\ge 2$
\begin{equation}
x_p^{(k)}=Ax_{p-1}^{(k)}+B
x_{p-2}^{(k)}\quad {\rm with}\,\,x_0^{(k)}=1\quad {\rm and}\,\, x_1^{(k)}\ {\rm given}. 
\end{equation}
Here $k=1,2$ labels the considered case,
namely $x_p^{(1)}=\|K^p\|$ and $x_p^{(2)}=\|K_p\widehat\Omega^0_{P,m}\|$. 
We know that $x_p$ is a linear combination of two geometric series
 $p\to r_\pm^p$. More explicitly $x_p=\alpha_+r_+^p+\alpha_-r_-^p$  where $r_\pm$
are the solutions of the characteristic equation 
$r^2-Ar-B=0$.
Denoting with $\Delta:=\sqrt{A^2+4B}$ one finds 
\begin{equation} 
r_\pm={A\pm\sqrt{\Delta}\over2}.
\end{equation} 
The coefficients of the linear combination $\alpha_\pm$ are solutions 
of the equation
\begin{equation}
\left (\begin{array}{cc}1&1\\
         r_+&r_-\end{array}\right )
\left (\begin{array}{c}\alpha_+\\ \alpha_-\end{array}\right )
=\left (\begin{array}{c}x_0\\ x_1\end{array}\right )\iff
\alpha_\pm=\pm{x_1-r_\mp x_0\over\sqrt{\Delta}}.
\end{equation} 
Now for the case 1 we have by direct calculation $\alpha_\pm^{(1)}={r_\pm\over \sqrt{\Delta}}$ 
and then 
$$
x_p^{(1)}={r_+^{p+1}-r_-^{p+1}\over\sqrt{\Delta}}
=\sum_{k=0}^pr_+^kr_-^{p-k}=
r_+^p\sum_{k=0}^p\left(r_-\over r_+\right)^{p-k}.
$$
For the second case we use the superposition principle.
 Let $y$  the sequence defined by the relation as $x$ 
with the initial conditions
$y_0=0$ et
$y_1= B$. Then we have 
$x^{(2)}=x^{(1)}+y$. Since $y_p=\beta_+r_+^p+\beta_-r_-^p$ with 
$\beta_\pm=\pm
B\Delta^{-{1\over2}}$ one has
$$
x_p^{(2)}=x_p^{(1)}+{B\over\sqrt{\Delta}}(r_+^p-r_-^p)=
x_p^{(1)}+B\sum_{k=0}^{p-1}r_+^kr_-^{p-k}.
$$
We proceed to the estimates by noticing that
$|r_-|< r_+$
which allows to write 
$$
\left|\sum_{k=0}^pr_+^kr_-^{p-k}\right|
=\left|r_+^p\sum_{k=0}^p\left(r_-\over
r_+\right)^{p-k}\right|\le r_+^p{1\over 1-{|r_-|\over r_+}}
={r_+^{p+1}\over
r_+-|r_-|}={r_+^{p+1}\over A}
$$
and consequently
\begin{equation}
|x_p^{(1)}|\le{ r_+^{p+1}\over A}\qquad
 |x_p^{(2)}|\le { r_+^{p+1}\over A}
+{B\over A} r_+^p.
\end{equation}
Using these results we obtain
\begin{equation}
\|K^p\id\|\le{r_+\over A}r_+^p\le 
\left(1+{\sqrt{B}\over A}\right)B^{p\over2}
\left(1+{A\over\sqrt{B}}\right)^p
\end{equation} 
where
$$
{\sqrt{B}\over A}
={1\over4\sqrt 3}\left(\log\left({a\over24}\vee 1\right)\right)^{1\over2}.
$$
The estimate on $x_p^{(2)}$ gives 
\begin{equation}
\|K^p{\widehat\Omega}_{P,m}^0\|\le { r_+^{p+1}\over A}+{B\over A} r_+^p
\leq \left (1+\frac{\sqrt B}{A}+\frac{B}{A}\right )B^{\frac{p}{2}}
\left (1+\frac{A}{\sqrt B} \right )
\end{equation} 
where
$$
{B\over A}=\frac{1}{4\sqrt 6}{|b|\over \sqrt{a}}\log({a\over 24}\vee1)
$$
Therefore we have proved:
\vsth
\noindent
{\sl {\bf Theorem 7.3.2 } 
Let ${\widehat {\cal U}}_{P,m}$ and  ${\widehat {\cal U}}_{P,m}^0$ as defined in Section 5. 
 $a=m-1$ and $b={\hat V}(2(m-1))$. Then there exists $m_0$ and a constant 
$C_{m_0}$ such that the Dyson series for 
${\widehat\Omega}_{P,m}^0(t,t_0)={\widehat {\cal U}}_{P,m}^0(t,t_0)^*
{\widehat {\cal U}}_{P,m}^0$ converges uniformly with respect to $t$ for every $t\in I_0$
and that for any $m>m_0$ one has the 
following estimate
\begin{equation}
\|{\widehat\Omega}_{P,m}^{0,(p)}\|_\infty \leq C_{m_0}|b|^p \sqrt {\log a} 
\left (\frac{ \log a}{a}\right )^{\frac{p}{2}}.
\end{equation}
\par
The remainder of order $p$ satisfy the estimate
\begin{equation}\label{reste}  
\|K^p {\widehat\Omega}_{P,m}^0\|_\infty
\leq 2C_{m_0}|b|^p \sqrt {\log a}
\left (\frac{ \log a}{a}\right )^{\frac{p}{2}}. 
\end{equation}}


\noindent {\bf Acknowledgments}
G. Nenciu and V. Moldoveanu would like to thank for the warm hospitality 
and financial support encountered along their visits at CPT Marseille and 
Universit\'e de Toulon et du Var. Parts of the draft were written
during the visits of F. Bentosela and P. Duclos at 
Institute of Mathematics of the Romanian Academy. The financial support 
from the Provence Alpes C\^{o}tes d'Azur region, Romanian Academy - CNRS (France)
exchange agreement, Grant CERES 38/02, Grant CNCSIS and Eurrommat Programme is
hereby acknowledged.

\end{document}